\documentclass[12pt]{article}

\usepackage{epsfig}				
\usepackage{amssymb,amsmath}
\usepackage{multirow}
\usepackage{rotating}
\usepackage{graphicx}
\usepackage{cite}				
\usepackage{lscape}
\usepackage{hyperref}
\usepackage{subcaption}			
\usepackage{color}
\usepackage{url}

\newcommand{\bea}{\begin{eqnarray}}
\newcommand{\eea}{\end{eqnarray}}

 \makeatletter
\def\alt{\mathrel{\mathpalette\gl@align<}}
\def\agt{\mathrel{\mathpalette\gl@align>}}
\def\gl@align#1#2{\lower.6ex\vbox{\baselineskip\z@skip\lineskip\z@
\ialign{$\m@th#1\hfil##\hfil$\crcr#2\crcr\sim\crcr}}} \makeatother

\begin{document}

\begin{flushright}
\end{flushright}

\vspace*{1.0cm}

\begin{center}
\baselineskip 20pt 
{\Large\bf 
Dark matter in  U(1) extensions of the MSSM  with gauge kinetic mixing
}
\vspace{1cm}

{\large 
Genevi\`eve B\'elanger$^{a,}$%
\footnote{ E-mail: belanger@lapth.cnrs.fr},
Jonathan Da Silva$^{b,}$%
\footnote{ E-mail: jonathan.da.silva.physics@gmail.com}
and 
Hieu Minh Tran$^{c,}$%
\footnote{ E-mail: hieu.tranminh@hust.edu.vn} 
} 
\vspace{.5cm}

{\baselineskip 20pt \it
$^a$LAPTH, Universite Savoie Mont Blanc, CNRS, B.P.110, F74941 Annecy Cedex, France \\
\vspace{2mm} 
$^b$Laboratoire de Physique Subatomique et de Cosmologie, Universit\'e Grenoble-Alpes,\\ 
CNRS/IN2P3, 53 Avenue Des Martyrs, F-38026 Grenoble, France\\
\vspace{2mm}
$^c$Hanoi University of Science and Technology, 
1 Dai Co Viet Road, Hanoi, Vietnam \\
}

\vspace{.5cm}

\vspace{1.5cm} {\bf Abstract}
\end{center}

The gauge kinetic mixing in general is allowed in models with multiple Abelian gauge groups.
In this paper, we investigate the gauge kinetic mixing in the framework of $U(1)$ extensions of the MSSM.
It enlarges the viable parameter space, 
and has an important effect on the particle mass spectrum 
as well as the $Z_2$ coupling with matters.
The SM-like Higgs boson mass can be enhanced with a nonzero kinetic mixing parameter and
the muon $g-2$ tension is 
slightly less severe than in the case of no mixing.
We present the results from both benchmark analysis and global parameter scan.
Various theoretical and phenomenological constraints
have been considered.
The recent LHC searches for the $Z_2$ boson are important for the case of large positive kinetic mixing where the $Z_2$ coupling is enhanced, 
and severely  constrain scenarios with 
$M_{Z_2} < 2.8$ TeV.
The viable dark matter candidate predicted by the model is either the neutralino or the right-handed sneutrino.
Cosmological constraints from dark matter searches play a significant role in excluding the parameter space.
Portions of the parameter space with relatively low sparticle mass spectrum can be successfully explored in the LHC run-2 as well as future linear colliders and dark matter searches.

\thispagestyle{empty}

\newpage

\addtocounter{page}{-1}

\baselineskip 18pt


\section{Introduction}  

Although the standard model (SM) has been verified to a very high accuracy, an extension is necessary both for theoretical consistency and in order to explain  experimental observations.
The minimal supersymmetric (SUSY) extension of the SM (MSSM) has played an important role in phenomenological studies for many years  because it could  address many fundamental issues such as the gauge hierarchy problem,
the prediction of the Higgs boson mass, and the gauge coupling unification while also providing  a dark matter (DM) candidate, the lightest neutralino.
Nevertheless, the discovery of a 125 GeV Higgs boson
\cite{Aad:2012tfa, Chatrchyan:2012xdj}
 has imposed some tension on the MSSM.
In order to reconcile the Higgs boson mass,  large loop corrections are needed, these in general require heavy squarks (especially stops) and a large mixing, thus reintroducing a certain amount of fine-tuning to the theory~\cite{Hall:2011aa}.
Moreover, the parameter regions with maximal stop mixing which allow to obtain the observed  Higgs mass  potentially have a metastable electroweak vacuum, 
and predict a global minimum which breaks charge and/or color symmetries 
\cite{Camargo-Molina:2013sta,Blinov:2013fta,Chowdhury:2013dka}.
In scenarios where the number of free parameters is limited due to some relations at a high energy scale, heavy squarks imply heavy sleptons 
\cite{Cao:2011sn,Okada:2012nr}.
Hence, the SUSY contribution to the muon anomalous magnetic dipole moment $(g-2)$ can hardly explain the discrepancy between the SM prediction and the experimental result.
This last issue is however easily resolved when allowing a larger hierarchy between the slepton and squark masses.


There have been attempts to resolve these tensions, see for example  
\cite{Okada:2016wlm,Yin:2016shg,Aboubrahim:2016xuz} and references therein. 
In this paper, we consider $U(1)'$ extensions of the MSSM (UMSSM) that can also improve the situation
\cite{Barger:2008jx,Belanger:2011rs,Athron:2012sq,Hirsch:2012kv,Belanger:2015cra}. 
The interaction between the extra singlet superfield, whose vacuum expectation value (VEV) breaks the $U(1)'$, and the two Higgs doublets helps to increase the mass of the SM-like  Higgs at the tree level.
This contribution is the same as in the next-to-MSSM (NMSSM)
\cite{Ellwanger:2014dfa,Kaminska:2014wia,Farina:2013fsa,Athron:2013ipa}.
In addition, the SM-like Higgs boson mass is also enhanced by the $U(1)'$ D-term contribution
\cite{Cvetic:1997ky,Barger:2006dh}.
Both these effects imply that the loop-induced contribution from the stop sector does not need to be large. 
In this framework, as in the NMSSM, the $\mu$-term problem is solved since this term is not introduced by hand but is generated by the vacuum expectation value of the singlet after the extra gauge group $U(1)'$ is broken.
The physics origin of the $U(1)'$ group depends on the specific scenario, for instance $U(1)_{B-L}$, or inherited from some grand unified theory (GUT).
Here we are interested in the scenario where the $U(1)'$ is a remnant symmetry after the breaking of the $E_6$ GUT
\cite{Langacker:1998tc}.
This scenario is also motivated by superstring models 
\cite{Cvetic:1995rj,Cvetic:1996mf}.

To account for the neutrino oscillations, we introduce in the model three generations of right-handed (RH) neutrinos. Assuming R-parity conservation,  their superpartners which are weakly interacting massive particles can play the role of  DM. This is in contrast with left-handed (LH) sneutrinos, which although also weakly interacting, have been ruled out as a DM candidate because 
their scattering cross section onto nuclei is too large~\cite{Falk:1994es}.
This model offers two possible candidates for the DM, the ordinary neutralino and the right handed sneutrino, depending on which one is the lightest superparticle (LSP).


A special feature of  models with two Abelian gauge groups $U(1) \times U(1)'$,  is  that a gauge kinetic mixing term can exist in the Lagrangian without violating any underlying symmetry 
\cite{Holdom:1985ag,Chankowski:2006jk,Brahmachari:2014aya,Babu:1997st}:
\begin{eqnarray}
\mathcal{L} & \supset & -\frac{k}{2} F^{\mu\nu} F'_{\nu\mu}.
\end{eqnarray}
Generally, even in the case that the kinetic mixing term is set to zero at some scale, it can be radiatively generated at the low energy scale due to the renormalization group (RG) evolution 
\cite{delAguila:1988jz,delAguila:1987st}.
It was found that in the $U(1)_{B-L}$ case the gauge kinetic mixing effect can be 
significant and impact DM observables~\cite{Fonseca:2011vn,O'Leary:2011yq,Basso:2012gz}. 

DM properties in U(1) extensions of the MSSM were examined in \cite{Kalinowski:2008iq,Belanger:2011rs,Athron:2016qqb,Athron:2016gor} and 
the compatibility of the UMSSM with collider and DM observables was examined in~\cite{Belanger:2015cra} where the kinetic mixing was neglected. Here we revisit and update the constraints on the parameter space of the UMSSM inspired from  $E_6$ GUT,  while including the kinetic mixing. 
The radiatively generated kinetic mixing term depends on the particle content and the charge assignment of fields under the two $U(1)$ gauge groups.
For example, in the minimal SUSY $B-L$ model \cite{O'Leary:2011yq} the kinetic mixing parameter purely induced from the RG evolution is positive and sizable, $k \sim \mathcal{O}(1)$,
while in $E_6$ models
\cite{Babu:1996vt} the value of k at low energies can be either positive or negative.
To be completely general, we will consider that 
$k$ is a free parameter  set at the low energy scale.

We will show that the gauge kinetic mixing can give rise to important effects on both the mass spectrum and DM properties. For example the kinetic mixing allows for  a leptophobic $Z_2$ which can more easily escape LHC constraints, gives a contribution to the  mass of the Higgs boson, can shift the mass of sleptons thus providing a better agreement with the muon anomalous magnetic moment, and finally impact the DM annihilation channel.
Note that in this study we include updated constraints from the LHC searches on a heavy neutral gauge boson $Z_2$ as well as updated constraints from DM direct detection from LUX~\cite{Akerib:2016vxi}.

The structure of the paper is as follows. 
In Section 2, we briefly describe the UMSSM model with gauge kinetic mixing. 
The effects of the kinetic mixing term on the parameter space, the $Z_2$ coupling with matter, and the mass spectrum  are shown in Section 3.
Here, benchmark analysis and results of the global parameter scan are presented with various collider constraints as well as cosmological ones taken into account.
Section 4 is devoted for conclusion.

\section{
The UMSSM with gauge kinetic mixing
}

\subsection{The model}

The UMSSM has the gauge groups $SU(3)_C \times SU(2)_L \times U(1)_Y \times U(1)'$ which remain after the symmetry breaking of an $E_6$ GUT.
The particle contents of this model include
the MSSM chiral supermultiplets, 
three generations of RH neutrino supermultiplets 
$N^c = \{ \tilde{\nu}_R^c, \nu_R^c \}$, 
the MSSM vector supermultiplets and
an additional vector supermultiplet $V' = \{ \tilde{B}' , B'_\mu \}$ corresponding to $U(1)'$ gauge group,
and a Higgs singlet superfield $S$ responsible for the $U(1)'$ breaking.
Additional chiral supermultiplets are included in an anomaly free $E_6$ theory. For simplicity, we assume that all the fields belong to the \textbf{27} representations of $E_6$ that are not listed above are heavy enough to be safely neglected at low energies.%

The $U(1)'$ charge of a chiral superfields is given by
\begin{eqnarray}
\mathcal{Q}' &=& \cos \theta_{E_6} \mathcal{Q}'_\chi + \sin \theta_{E_6} \mathcal{Q}'_\psi \, ,
\end{eqnarray}
where $\theta_{E_6} \in [-\frac{\pi}{2},\frac{\pi}{2}]$ parameterizes a linear combination of two $E_6$ subgroups $U(1)_\chi$ and $U(1)_\psi$ into $U(1)'$.
The charges $\mathcal{Q}'_\chi$ and $\mathcal{Q}'_\psi$ for each chiral superfield of the model are given in Table \ref{charges}.

\begin{table}
\begin{center}
\begin{math}
\begin{array}{|c|ccccccccc|}
\hline
	& Q	& U^c	& D^c	& L	& N^c	& E^c	& H_u	& H_d	& S	\\
\hline
\sqrt{40} \mathcal{Q}'_\chi	& -1	& -1	& 3	& 3	& -5	& -1	&  2	& -2	& 0	\\
\sqrt{24} \mathcal{Q}'_\psi	&  1 	&  1	& 1	& 1	&  1	&  1	& -2	& -2	& 4	\\
\hline
\end{array}
\end{math}
\caption{$U(1)'$ charges of chiral superfields.}
\label{charges}
\end{center}
\end{table}


The superpotential of the model involves the ordinary MSSM superpotential without the $\mu$-term, and other terms describing interactions of the Higgs singlet and right handed neutrinos:
\begin{eqnarray}
\mathcal{W} &\supset& 
	\mathcal{W}_\text{MSSM}|_{\mu=0} + \lambda S H_u H_d + N^c \mathbf{Y_\nu} L H_u \, ,
\end{eqnarray}
where $\mathbf{Y}_\nu$ is the neutrino Yukawa coupling matrix responsible for the neutrino mass generation.
After the $U(1)'$ group is broken, the $\mu$-term is generated by the singlet's VEV,
$\langle S \rangle = \frac{v_S}{\sqrt{2}}$ as
\begin{eqnarray}
\mu &=& \lambda \frac{v_S}{\sqrt{2}} \, .
\label{mu}
\end{eqnarray}

The soft SUSY breaking Lagrangian of the UMSSM reads
\begin{eqnarray}
\mathcal{L}^\text{soft} &\supset&
	\mathcal{L}^\text{soft}_\text{MSSM}|_{B_\mu=0} 
	- \left( 
	\frac{1}{2} M'_1 \tilde{B}' \tilde{B}' 
	+ \tilde{\nu}^c_R \mathbf{A_\nu} \tilde{L} H_u + h.c.
	\right)
	- \tilde{\nu}^c_R \mathbf{M^2_{\tilde{\nu}_R}} \tilde{\nu}_R \nonumber \\
	&&
	- m_S^2 |S|^2
	- (\lambda A_\lambda S H_u H_d + h.c.) \, ,
\end{eqnarray}
where new soft terms are added in comparison to the MSSM: 
the $\tilde{B}'$ soft mass, $M_1'$,
the neutrino trilinear couplings, $\mathbf{A_\nu}$,
the right handed sneutrinos soft masses, $\mathbf{M^2_{\tilde{\nu}_R}}$,
the singlino mass, $m_S$,
and the Higgs trilinear coupling, $A_\lambda$.
Similar to Eq. (\ref{mu}), the MSSM $B\mu$ term is induced by the $U(1)'$ breaking:
\begin{eqnarray}
B\mu &=& \lambda A_\lambda \frac{v_S}{\sqrt{2}} \, .
\end{eqnarray}

\subsection{Gauge kinetic mixing}

The general gauge kinetic Lagrangian for Abelian gauge superfields is written as follows
\begin{eqnarray}
\mathcal{L}_\text{kinetic}^\text{gauge} &\supset&
	- \int d^4 \theta
	\frac{1}{4}
	\left(
		\begin{array}{cc}
		W^\alpha & W'^\alpha
		\end{array}
	\right)
	\left(
		\begin{array}{cc}
		1	&	k	\\
		k	&	1
		\end{array}
	\right)
	\left(
		\begin{array}{c}
		W_\alpha \\ W'_\alpha
		\end{array}
	\right)
	+ h.c. \, ,
\label{kineticL}
\end{eqnarray}
where the off-diagonal element $k$ is the gauge kinetic mixing parameter.
The kinetic mixing matrix can be diagonalized by a rotation among the original Abelian vector superfields, ($\hat{V},\hat{V}'$):
\begin{eqnarray}
\left(
	\begin{array}{c}
	\hat{V}	\\	\hat{V}'
	\end{array}
\right) &=&
	\left(
		\begin{array}{cc}
		\frac{1}{\sqrt{2(1+k)}}	&	\frac{-1}{\sqrt{2(1-k)}}	\\
		\frac{1}{\sqrt{2(1+k)}}	&	\frac{1}{\sqrt{2(1-k)}}
		\end{array}
	\right)	
	\left(
		\begin{array}{c}
		V_Y	\\	V_E
		\end{array}
	\right) \, .
\label{rotation1}
\end{eqnarray}
For a real rotation, the kinetic mixing parameter is limited to 
$-1 < k < 1$.
The rotation (\ref{rotation1}) ensures that there is no explicit kinetic mixing in the Lagrangian written in the new basis ($V_Y,V_E$).
However, the effect of the kinetic mixing term now transfers to the interactions between the Abelian vector superfields and chiral superfields.
The gauge interaction Lagrangian is
\begin{eqnarray}
\mathcal{L}^\text{gauge}_\text{interaction} & \supset &
	\int d^4 \theta
	\Phi^\dagger e^{\textbf{Q} \cdot \textbf{g} \cdot\textbf{V}} \Phi	\, ,
\end{eqnarray}
where
\begin{eqnarray}
\textbf{Q} \cdot \textbf{g} \cdot\textbf{V} 
&=&	\left(
		\begin{array}{cc}
		Y	&	\mathcal{Q}'
		\end{array}
	\right)
	\left(
		\begin{array}{cc}
		g_{YY}	&	g_{YE}	\\
		g_{EY}	&	g_{EE}
		\end{array}
	\right)
	\left(
		\begin{array}{c}
		V_Y	\\	V_E
		\end{array}
	\right)	\, .
\label{gauge}
\end{eqnarray}
where $Y$ is the hypercharge and $\mathcal{Q}'$ the charge associated with $U(1)'$.
The gauge coupling matrix which is originally diagonal absorbs the rotation of the Abelian vector superfields, and becomes non-diagonal:
\footnote{
In our analysis in the next section, we will assume for simplicity that $g_1' = \sqrt{\frac{5}{3}} g_1.$
}
\begin{eqnarray}
\left(
	\begin{array}{cc}
	g_{YY}	&	g_{YE}	\\
	g_{EY}	&	g_{EE}
	\end{array}
\right)		
&=&	\left(
		\begin{array}{cc}
		\frac{g_1}{\sqrt{2(1+k)}}	&	\frac{-g_1}{\sqrt{2(1-k)}}	\\
		\frac{g'_1}{\sqrt{2(1+k)}}	&	\frac{g'_1}{\sqrt{2(1-k)}}
		\end{array}
	\right)	\, .
\end{eqnarray}

To simplify the gauge coupling matrix, we perform an orthogonal rotation in the space of Abelian vector superfields such that the gauge kinetic matrix remains intact:
\begin{eqnarray}
\left(
	\begin{array}{c}
	V_Y	\\	V_E
	\end{array}
\right) &=&
	\frac{1}{\sqrt{g_{EE}^2 + g_{EY}^2}}
	\left(
		\begin{array}{cc}
		g_{EE}	&	g_{EY}	\\
		-g_{EY}	&	g_{EE}
		\end{array}
	\right)	
	\left(
		\begin{array}{c}
		V	\\	V'
		\end{array}
	\right)	\, .
\end{eqnarray}
Eq. (\ref{gauge}) is then rewritten as
\begin{eqnarray}
\textbf{Q} \cdot \textbf{g} \cdot\textbf{V} 
&=&	\left(
		\begin{array}{cc}
		Y	&	\mathcal{Q}'
		\end{array}
	\right)
	\left(
		\begin{array}{cc}
		g_{y}	&	g'	\\
		0	&	g_{E}
		\end{array}
	\right)
	\left(
		\begin{array}{c}
		V	\\	V'
		\end{array}
	\right)	\, ,
	\label{eq:rotation}
\end{eqnarray}
in which
\begin{eqnarray}
g_y =	&	\dfrac{g_{YY} g_{EE} - g_{YE} g_{EY}}{\sqrt{g_{EE}^2 + g_{EY}^2}}	&
		=	g_1		\,	,	\\
g'	=	&	\dfrac{g_{YY} g_{EY} + g_{YE} g_{EE}}{\sqrt{g_{EE}^2 + g_{EY}^2}}	&
		=	\frac{-k g_1}{\sqrt{1-k^2}}	\,	,	\\
g_E	=	&	\sqrt{g_{EE}^2 + g_{EY}^2} 	&
		=	\frac{g'_1}{\sqrt{1-k^2}}	\, .
\label{gE-define}
\end{eqnarray}
Note that in the limit $k \rightarrow 0$, the above Abelian gauge coupling matrix becomes diagonal.
Performing matrix multiplication in Eq.~\ref{eq:rotation}, we obtain:
\begin{eqnarray}
\textbf{Q} \cdot \textbf{g} \cdot\textbf{V} 
&=&
		Y g_1 V + Q^p g_E V'	\,	,
\label{redefinition}
\end{eqnarray}
where 
the new charge $Q^p$ is defined as
\begin{eqnarray}
Q^p 
&=&	\mathcal{Q}' - k \frac{g_1}{g'_1} Y	\,	.
\label{Qp}
\end{eqnarray}
Clearly, $Y$ corresponds to the SM hypercharge and $V$ is the associated gauge superfield while the kinetic mixing induces a shift  in the new charge of the chiral superfields, from $\mathcal{Q}' \to Q^p$,
and the coupling with the new Abelian superfield,
from $g_1' \to g_E$.
It is worth to note that the anomaly cancellation conditions for $\{\mathcal{Q}', Y\}$ in the underlying $E_6$ theory ensure the theory to be anomaly free for the redefined charge $Q^p$.

\subsection{Neutral gauge bosons}

The original Abelian vector superfields $(\hat{V},\hat{V}')$ are mixed to form the new ones $(V,V')$.
Their vector components $(B_\mu, B'_\mu)$ in turn mix with the third component $W^3_\mu$ of the $SU(2)_L$ gauge group to form mass eigenstates $(A_\mu, Z_{1\mu}, Z_{2\mu})$ when the gauge groups $SU(2)_L \times U(1)_Y \times U(1)'$ are broken spontaneously.
The $Z$-boson mixing mass matrix is as follows
\begin{eqnarray}
\mathbf{M_{Z}^2} &=&
		\left(
		\begin{array}{cc}
		M_{ZZ}^2	&	M_{ZZ'}^2	\\
		M_{ZZ'}^2	&	M_{Z'Z'}^2	
		\end{array}
	\right)	\, ,
\end{eqnarray}
where
\begin{eqnarray}
M_{ZZ}^2 	&=&	\frac{1}{4} g_1^2 (v_u^2 + v_d^2) \, ,	\nonumber	\\
M_{Z'Z'}^2 	&=&	g_E^2 
	\left[ (Q_{H_u}^p)^2 v_u^2 + (Q_{H_d}^p)^2 v_d^2
			+ (Q_S^{p})^2 v_S^2
	\right] \, ,	
	\label{mz2}\\
M_{ZZ'}^2	&=&	\frac{1}{2} g_1 g_E 
				( Q_{H_u}^p v_u^2 - Q_{H_d}^p v_d^2 ). \nonumber
\end{eqnarray}
This matrix can be diagonalized by an orthogonal rotation:
\begin{eqnarray}
\left( \begin{array}{c}
		Z_1		\\
		Z_2	
		\end{array}
\right) &=&
	\left(
	\begin{array}{cc}
	\cos \alpha_Z	&	\sin \alpha_Z	\\
	-\sin \alpha_Z	&	\cos \alpha_Z
	\end{array}
	\right)
	\left( \begin{array}{c}
		Z^0		\\
		Z'	
		\end{array}
	\right) \, ,
\label{mzz}
\end{eqnarray}
where $\alpha_Z$ is the mixing angle defined as
\begin{eqnarray}
\sin 2\alpha_Z &=& \frac{2M_{ZZ'}^2}{M_{Z_2}^2-M_{Z_1}^2}
\label{az}
\end{eqnarray}
The physical states $Z_1$ and $Z_2$ have masses:
\begin{eqnarray}
M^2_{Z_1,Z_2} &=&
	\frac{1}{2} 
	\left[
	M^2_{ZZ} + M^2_{Z'Z'} \mp 
	\sqrt{ \left( M^2_{ZZ} - M^2_{Z'Z'}\right)^2 + 4 M^4_{ZZ'}}
	\right]
\end{eqnarray}
In our analysis, we use the measured Z-boson mass for $M_{Z_1}$, while $M_{Z_2}$ and $\alpha_Z$ are considered as free parameters.

\subsection{Sfermions and neutralinos}

In the UMSSM, the D-term contributions to sfermion masses play an important role in forming the sparticle mass spectrum.
They modify the diagonal components of the usual MSSM sfermion mass matrices as
\begin{eqnarray}
\Delta_{\tilde{f}} &=&
	\frac{1}{2} g_E^2 Q^p_{\tilde{f}}
	\left(
	Q^p_{H_u} v_u^2 + Q^p_{H_d} v_d^2 + Q^p_S v_S^2
	\right)	\, ,
\label{D-term}
\end{eqnarray}
where
$\tilde{f} = \{ \tilde{q}_L^i, \tilde{u}_R^i, \tilde{d}_R^i, \tilde{l}_L^i, \tilde{\nu}_R^i, \tilde{e}_R^i \}$ with the generation index $i=\{1,2,3\}$.
Sine the redefined charges $Q^p$ and gauge coupling $g_E$ are functions of $k$, the sparticle mass spectrum also depends on the kinetic mixing parameter.
As we will see, this effect is particularly important.

While charginos are the same as in the MSSM, the neutralino sector of the UMSSM consists of six fermions.
Their masses are eigenvalues obtained from the mass matrix that is written in the basis of neutral fermionic components of the vector supermultiplets and the Higgs supermultiplets 
$\psi^0 = (\tilde{B}, \tilde{W}^3, \tilde{H}_d, \tilde{H}_u, \tilde{S}, \tilde{B}')^T$ as
\begin{eqnarray}
\mathbf{M_{\tilde{\chi}^0}} &=&
\left(
	\begin{array}{cccccc}
	M_1	&	0	&	-M_{ZZ}c_\beta s_W	&	M_{ZZ}s_\beta s_W	&	0	&	0	\\
	0	&	M_2	&	M_{ZZ}c_\beta c_W	&	-M_{ZZ}s_\beta c_W	&	0	&	0	\\
	-M_{ZZ}c_\beta s_W	&	M_{ZZ}c_\beta c_W	&	0	&	-\mu	&	-\lambda \frac{v_u}{\sqrt{2}}	&	Q^p_{H_d}g_E v_d	\\
	M_{ZZ}s_\beta s_W	&	-M_{ZZ}s_\beta c_W	&	-\mu&	0	&	-\lambda \frac{v_d}{\sqrt{2}}	&	Q^p_{H_u}g_E v_u	\\
	0	&	0	&	-\lambda \frac{v_u}{\sqrt{2}}	&	-\lambda \frac{v_d}{\sqrt{2}}	&	0	&	Q^p_S g_E v_S	\\
	0	&	0	&	Q^p_{H_d}g_E v_d	&	Q^p_{H_u}g_E v_u&	Q^p_S g_E v_S	&	M'_1
	\end{array}
\right)	\, ,\nonumber	\\
\end{eqnarray}
where 
$c_W = \cos \theta_W , \, s_W = \sin \theta_W$,
$c_\beta = \cos \beta, \, s_\beta = \sin \beta$, 
with $\tan \beta = \frac{v_u}{v_d}$.
The value of $\tan\beta$ can be derived using Eqs. (\ref{mzz}) and (\ref{az}). We have
\begin{eqnarray}
\cos^2 \beta &=&
	\frac{1}{Q^p_{H_u} + Q^p_{H_d}}
	\left(
		\frac{\sin 2\alpha_Z \, (M_{Z_1}^2 - M_{Z_2}^2)}{v^2 g_E \sqrt{g_1^2 + g_2^2}} + 
		Q^p_{H_u}
	\right)	\, ,
\label{cosbeta}
\end{eqnarray}
in which $v^2 = v_u^2 + v_d^2$ and $g_2$ is the SU(2) coupling.
The matrix $\mathbf{Z_n}$ diagonalizing the above mass matrix determines the components of each neutralino:
\begin{eqnarray}
\tilde{\chi}^0_i &=& (\mathbf{Z_n})_{ij} \psi^0_j	\, ,
	\qquad	i,j = \{1,2,3,4,5,6 \}	\, ,
\end{eqnarray}
and therefore its properties.

\subsection{Higgs sector}

The tree level mass-squared matrix of CP-even Higgs bosons is a symmetric $3 \times 3$ matrix $\mathcal{M^0_+}$ with elements computed as:
\begin{eqnarray}
(\mathcal{M^0_+})_{11} &=&
	\left[
	\frac{g_1^2 + g_2^2}{4} + (Q^p_{H_d})^2 g_E^2
	\right] v_d^2 +
	\frac{\lambda A_\lambda v_S v_u}{\sqrt{2} v_d}	\, ,
	\nonumber	\\
(\mathcal{M^0_+})_{12} &=&
	-\left[
	\frac{g_1^2 + g_2^2}{4} -\lambda^2 - 
		Q^p_{H_u} Q^p_{H_d} g_E^2
	\right] v_u v_d -
	\frac{\lambda A_\lambda v_S}{\sqrt{2}}	\, ,
	\nonumber	\\
(\mathcal{M^0_+})_{13} &=&
	\left[
	\lambda^2 + Q^p_{H_d} Q^p_S g_E^2
	\right] v_S v_d
	-\frac{\lambda A_\lambda v_u}{\sqrt{2}}	\, ,	\\
(\mathcal{M^0_+})_{22} &=&
	\left[
	\frac{g_1^2 + g_2^2}{4} + (Q^p_{H_u})^2 g_E^2
	\right] v_u^2 +
	\frac{\lambda A_\lambda v_S v_d}{\sqrt{2} v_u}	\, ,
	\nonumber	\\
(\mathcal{M^0_+})_{23} &=&
	\left[
	\lambda^2 + Q^p_{H_u} Q^p_S g_E^2
	\right] v_S v_u
	-\frac{\lambda A_\lambda v_d}{\sqrt{2}}	\, ,
	\nonumber	\\
(\mathcal{M^0_+})_{33} &=&
	(Q^p_S)^2 g_E^2 v_S^2 +
	\frac{\lambda A_\lambda v_u v_d}{\sqrt{2} v_S}	\, .
	\nonumber
\end{eqnarray}
The lightest Higgs boson is the SM-like one. 
Its tree level mass can be written approximately as \cite{King:2005jy}
\begin{eqnarray}
m_{h_1}^2|_\text{tree}	&\simeq&
	M_{ZZ}^2 \cos^2 2\beta	+
	\frac{1}{2} \lambda^2 v^2 \sin^2 2\beta +
	g_E^2 v^2 (Q^p_{H_d} \cos^2 \beta + Q^p_{H_u}\sin^2 \beta )^2		\nonumber	\\
&&	- \frac{\lambda^4 v^2}{g_E^2 (Q_S^p)^2}	
		\left[
		1 - \frac{A_\lambda \sin^2 2\beta}{2\mu} +
		\frac{g_E^2}{\lambda^2} 
		(Q^p_{H_d} \cos^2 \beta + Q^p_{H_u}\sin^2 \beta)Q_S^p
		\right]^2	\,	.
\label{mh1}
\end{eqnarray}
While the second term in the above equation is the same as in the NMSSM, the last two terms only appear in the UMSSM due to the existence of $Q^p$ and $g_E$ related to the extra $U(1)'$.
Therefore the Higgs boson mass depends on the kinetic mixing parameter via these terms. Similarly the masses of $h_2$ and $h_3$ can receive large corrections due to the kinetic mixing.
There is one CP-odd Higgs $A^0$ with the mass:
\begin{eqnarray}
m_{A^0}^2|_\text{tree}	&=&
	\frac{\lambda A_\lambda \sqrt{2} }{\sin 2\beta} v_S
	\left(
	1 + \frac{v^2}{4 v_S^2} \sin^2 2\beta
	\right)	\,	.
\end{eqnarray}
The mass of the charged Higgs bosons is given by
\begin{eqnarray}
m_{H^\pm}^2|\text{tree} &=&
	M_W^2 + \frac{\sqrt{2} \lambda A_\lambda}{\sin 2\beta}v_S
	-\frac{\lambda^2}{2} v^2 \, .
\end{eqnarray}
These masses also depend on $k$ through the angle $\beta$, see Eq.~\ref{cosbeta}.

\section{Analysis}

\subsection{Theoretical constraints}

In Eq. (\ref{redefinition}), we have interpreted $g_E$ as a redefined $U(1)'$ gauge coupling.
It is crucial to check under which condition this new coupling satisfies the perturbation limit:
\begin{eqnarray}
\alpha_E = \frac{g_E(k)^2}{4\pi} \lesssim 1	\, .
\label{aE}
\end{eqnarray}
Replacing  with the  coupling definition in Eq.~\ref{gE-define}, this condition leads to an upper bound on $k^2<1- g_1'^2/4\pi$. This constraint is weak and  only excludes the regions of $k$ close to $\pm 1$.

\begin{figure}
\centering
\begin{subfigure}{0.5\textwidth}
  \centering
  \includegraphics[width=8.2cm]{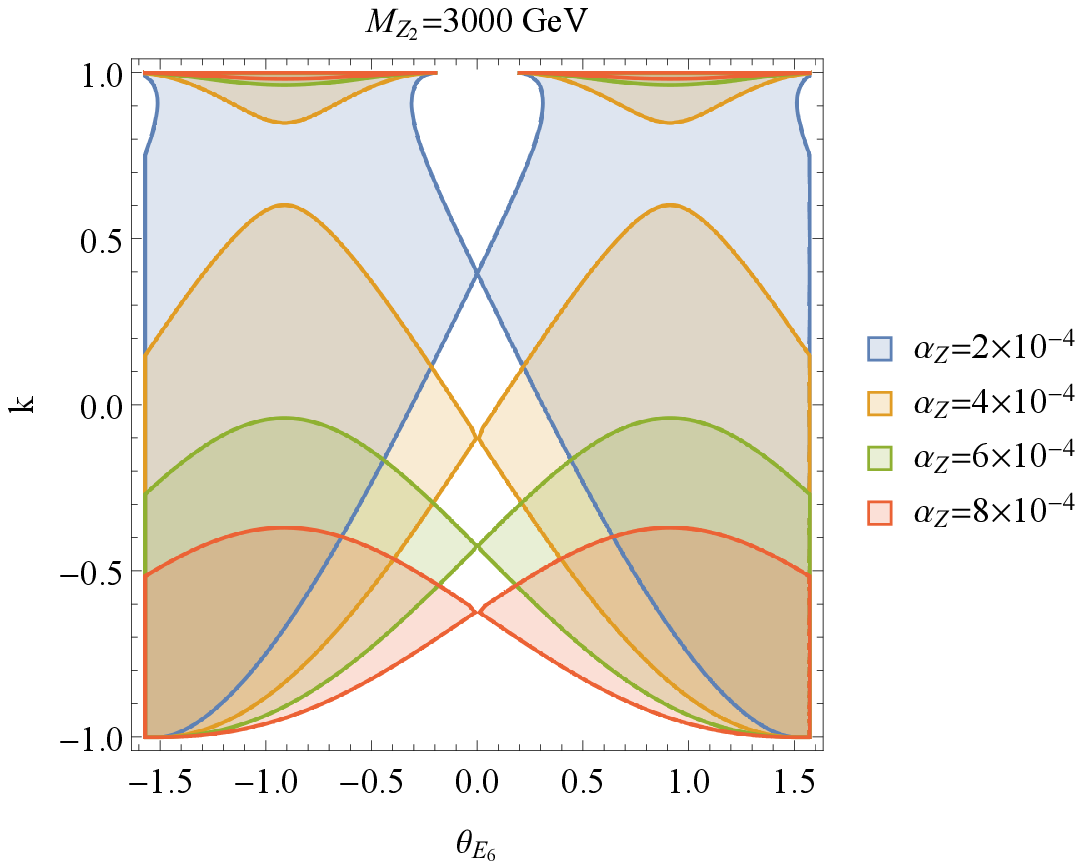}
  \caption{}
  \label{fig:sub1}
\end{subfigure}%
\begin{subfigure}{0.5\textwidth}
  \centering
  \includegraphics[width=8.2cm]{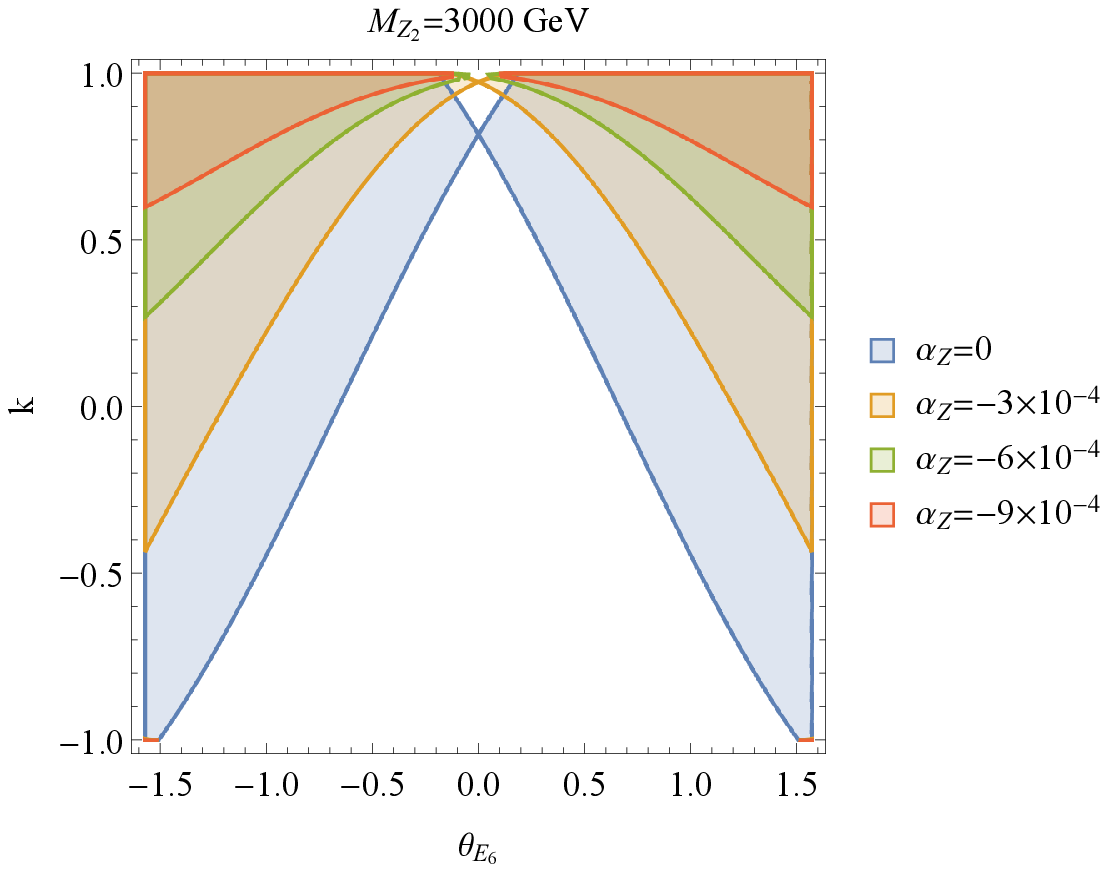}
  \caption{}
  \label{fig:sub2}
\end{subfigure}
\caption{Allowed regions in the $(\theta_{E_6},k)$ plane for 
the case of the $Z_2$ boson mass $M_{Z_2}=3000$ GeV and various values of the angle $\alpha_Z$:
(\ref{fig:sub1}) $\alpha_Z > 0$, 
(\ref{fig:sub2}) $\alpha_Z \leq 0$. } 
\label{cosb1}
\end{figure}

\begin{figure}
\begin{subfigure}{0.5\textwidth}
  \centering
  \includegraphics[width=8.2cm]{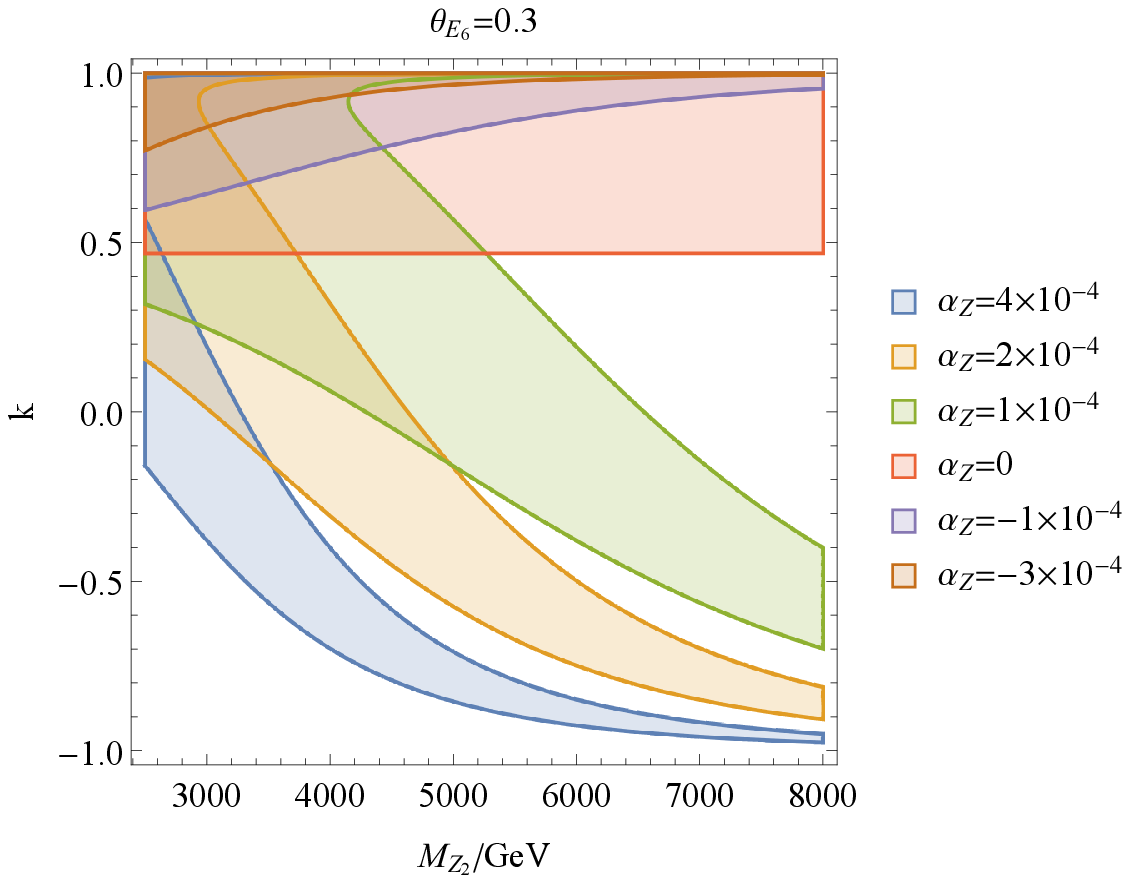}
  \caption{}
  \label{fig:sub3}
\end{subfigure}%
\begin{subfigure}{0.5\textwidth}
  \centering
  \includegraphics[width=8.2cm]{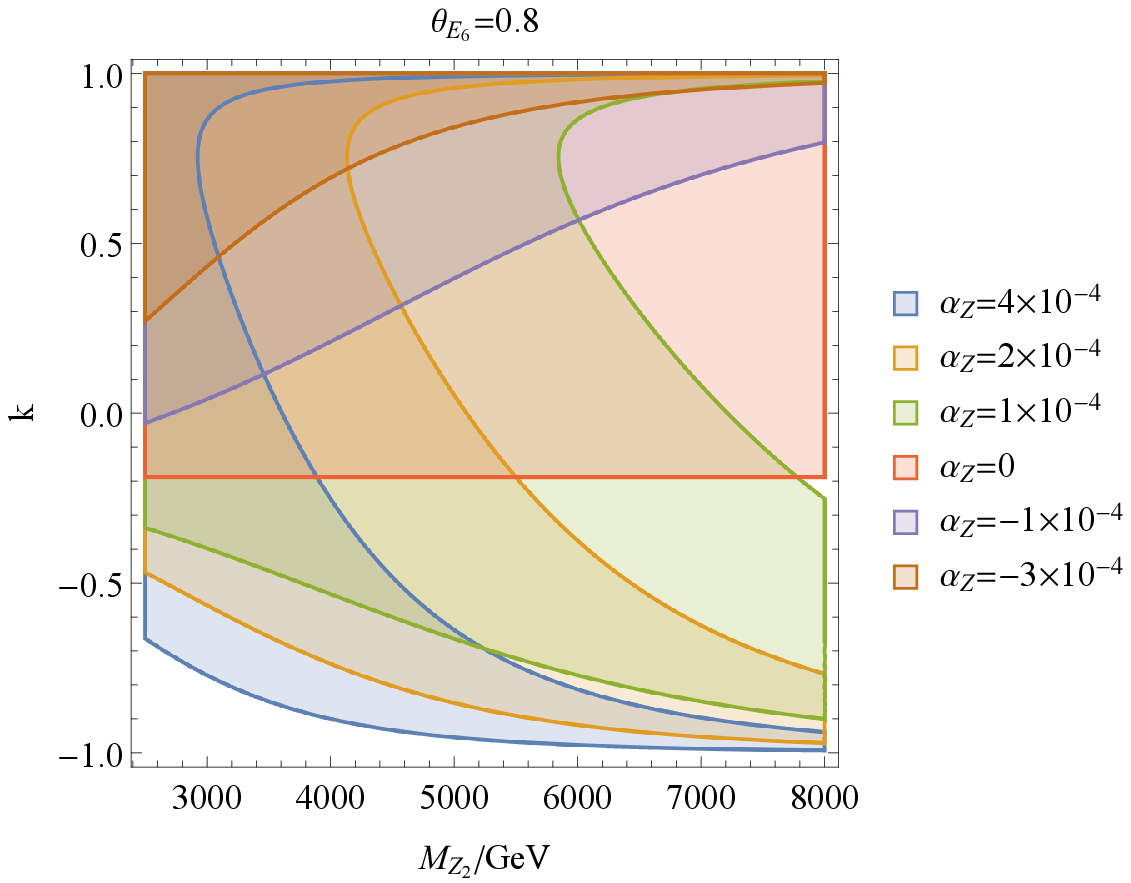}
  \caption{}
  \label{fig:sub4}
\end{subfigure}
\caption{Allowed regions in the $(M_{Z_2},k)$ plane for various values of $\alpha_Z$. The angle $\theta_{E_6}$ is set to 0.3 and 0.8 in (\ref{fig:sub3}) and (\ref{fig:sub4}) respectively. } 
\label{cosb2}
\end{figure}

\begin{figure}
\begin{subfigure}{0.5\textwidth}
  \centering
  \includegraphics[width=7.8cm]{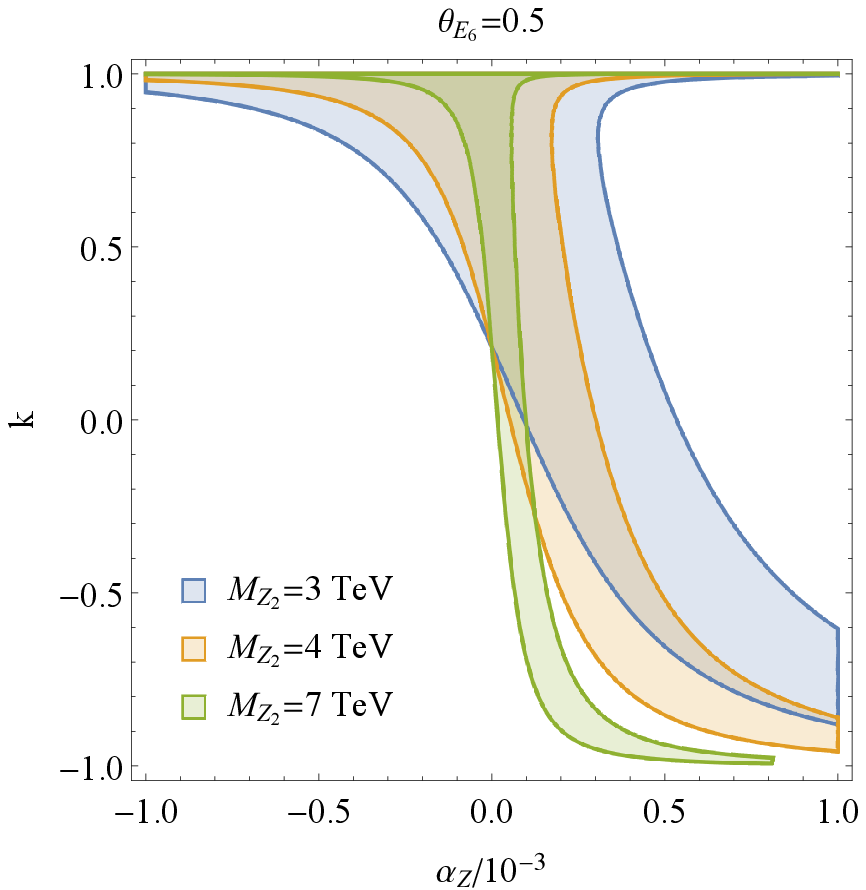}
  \caption{}
  \label{fig:sub5}
\end{subfigure}%
\begin{subfigure}{0.5\textwidth}
  \centering
  \includegraphics[width=7.8cm]{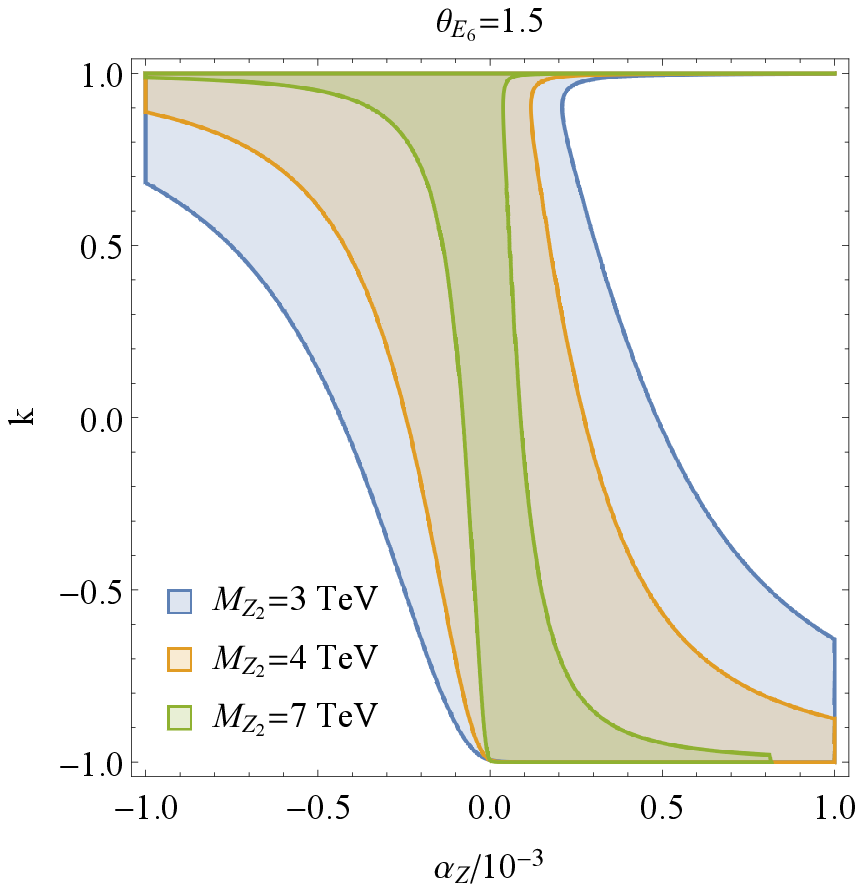}
  \caption{}
  \label{fig:sub6}
\end{subfigure}
\caption{Allowed regions in the $(\alpha_Z,k)$ plane for various values of the $Z_2$ boson mass $M_{Z_2}$.
The angle $\theta_{E_6}$ is chosen to be 0.5 and 1.5 in (\ref{fig:sub5}) and (\ref{fig:sub6}) respectively.}
\label{cosb3}
\end{figure}

In this model, $\tan\beta$ is not chosen as an independent parameter as in the MSSM.
It depends on the values of four other free parameters $M_{Z_2}$, $\alpha_Z$, $\theta_{E_6}$, and the kinetic mixing $k$ as expressed in Eq. (\ref{cosbeta}).
The reality condition on the angle $\beta$,
\begin{eqnarray}
0 \leq \cos^2 \beta \leq 1	\,	,
\label{consb}
\end{eqnarray}
defines the regions in  the parameter space of
$\{ M_{Z_2}, \alpha_Z, \theta_{E_6}, k \}$ where
further calculations can be carried out.
In Figs \ref{cosb1}, \ref{cosb2} and \ref{cosb3}, we show the parameter regions allowed by the constraint (\ref{consb}).
For a specific choice of $\{\theta_{E_6},\alpha_Z,M_{Z_2}\}$, the kinetic mixing parameter $k$ is limited to a specific range that is usually smaller than the open range $(-1,1)$.
Thus by allowing a nonzero kinetic mixing term, the acceptable ranges for other parameters change significantly.

First note that $Q^p_{H_u}+Q^p_{H_d}=-\frac{4}{\sqrt{24}}\sin\theta_{E_6}$. Thus at $\theta_{E_6}=0$, there is a unique value of $k$ that satisfies Eq.~\ref{cosbeta} for each choice of $M_{Z'}$ and $\alpha_Z$. This can be seen in 
Fig. \ref{cosb1} where  the allowed parameter regions in the plane $(\theta_{E_6},k)$ for the case of $M_{Z_2}=3000$ GeV are depicted. 
Moreover this value of $k$ is large and positive for $\alpha_Z<0$, Fig. \ref{fig:sub2}.  
Given a choice of $(M_{Z_2},\alpha_Z)$, for larger values of $\theta_{E_6}$ the range of allowed values for $k$ increases. 
The sign of $k$ is generally anticorrelated with that of $\alpha_Z$ for large values of the mixing to allow for a cancellation between the 
two terms in Eq.~\ref{cosbeta}, except when $k \sim +1$. 
Moreover $|k|$ approaches  1 as the $Z-Z'$ mixing increases.
Note that  for all cases where the first term in Eq.~\ref{cosbeta} dominates,  the allowed regions are symmetric with respect to a sign flip of $\theta_{E_6}$.

In Fig. \ref{cosb2}, we plot the allowed regions in the plane $(M_{Z_2},k)$ for various values of $\alpha_Z$ and two choices of $\theta_{E_6}$.
In the limit of no $Z-Z'$ mixing, $\alpha_Z\approx 0$, the range of values of $k$ become independent of $M_{Z_2}$ and are only set by the conditions $Q^p_{H_d},Q^p_{H_u} < 0$ for $\theta_{E_6} > 0$, and $Q^p_{H_d},Q^p_{H_u} > 0$ for $\theta_{E_6} < 0$.
Thus the non-zero kinetic mixing implies that regions of  parameter space with small values of  $\theta_{E_6}$ and small mixing $\alpha_Z$ are accessible while they were not with $k=0$~\cite{Belanger:2015cra}. However phenomenological constraints that will be discussed in the next section further restrict this region.
For non-zero mixing angles, $\alpha_Z$,  larger values of $|k|$ are required to increase $g_E$ and compensate an increase in $M_{Z_2}$  in the first term in Eq.~\ref{cosbeta}. 
Note that the allowed range for $k$ is quite narrow at large values of $M_{Z_2}$ and that
the allowed regions in the  plane $(M_{Z_2},k)$  become much larger for $\theta_{E_6}=0.8$  than for $\theta_{E_6}=0.3$.

We also show in
Fig. \ref{cosb3}  the allowed regions in the plane 
$(\alpha_Z,k)$ for various values of the $Z_2$ boson mass $M_{Z_2}$.
Figs. \ref{fig:sub5} and \ref{fig:sub6} correspond to $\theta_{E_6}= 0.5$ and 1.5 respectively.
For $\theta_{E_6}= 0.5$, only a narrow range of $Z-Z'$ mixing angles are allowed for $k\approx 0$, while for $k\approx 1$ any value is allowed. Indeed in this case  the first term in Eq.~\ref{cosbeta} 
becomes strongly suppressed.  As mentioned above,  for $\theta_{E_6}\approx \pi/2$, a larger area of parameter space is theoretically consistent.

In summary, the presence of the kinetic mixing enlarges significantly the theoretically allowed regions of parameter space, 
in particular  regions with small values of $\theta_{E_6}$, large mixing $\alpha_Z$ and low $Z_2$ boson mass.  
Moreover the  large positive kinetic mixing ($k \lesssim +1$) is slightly  favoured as compared to  large negative  ($k \gtrsim -1$)  as shown in Figs. \ref{cosb1}, \ref{cosb2} and \ref{cosb3}.

Besides the above theoretical constraints, we also impose perturbative Yukawa couplings, for this we require
 the Yukawa couplings  to be smaller than $\sqrt{4\pi}$ at the SUSY scale.
This constraint excludes the possibilities of very small or large values of $\tan \beta$.
We also require that the width to mass ratios of Higgs particles should satisfy
$\Gamma_{h_i}/m_{h_i} < 1$.

\subsection{Phenomenological constraints}

In our analysis, various phenomenological constraints are taken into account.
For the Higgs boson mass, the combined result of the ATLAS and CMS measurements is employed 
\cite{Aad:2015zhl} with a theoretical uncertainty of about 2 GeV.
The deflection $\Delta\rho$ of the electroweak $\rho$-parameter with respect to 1 is computed and compared to current upper bound \cite{Agashe:2014kda}.
We also consider the constraint on the muon anomalous magnetic moment $\Delta a_\mu$ 
\cite{Bennett:2006fi,Roberts:2010cj,Aoyama:2012wk}.
A variety of constraints from flavor physics are taken into account.
Observables in the $B$-meson sector that are of interest include:
the oscillation parameters
$\Delta M_s$, 
$\Delta M_d$ \cite{Amhis:2014hma},
the branching ratios of the following processes:
$B^\pm \rightarrow \tau^\pm \nu_\tau$ \cite{b_taunu},
$\bar{B}^0 \rightarrow X_s \gamma$ \cite{b_sg},
$B^0_s \rightarrow \mu^+ \mu^-$ \cite{Agashe:2014kda},
$\bar{B}^0 \rightarrow X_s \ell^+ \ell^-$ at low and high dilepton invariant mass \cite{Huber:2015sra},
$b \rightarrow d \gamma$ \cite{delAmoSanchez:2010ae,Crivellin:2011ba},
$B^0_d \rightarrow \mu^+ \mu^-$ \cite{Amhis:2014hma},
$B \rightarrow X_s \nu \bar{\nu}$ \cite{Barate:2000rc},
$B^+ \rightarrow K^+ \nu \bar{\nu}$ \cite{Amhis:2014hma},
$B^0 \rightarrow K^{*0} \nu \bar{\nu}$ \cite{Amhis:2014hma},
and the ratios
$R_D = \frac{\text{BR}(B^+ \rightarrow D \tau^+ \nu_\tau)}{\text{BR}(B^+ \rightarrow D \ell^+ \nu_\ell)}$,
$R_{D^*} = \frac{\text{BR}(B^+ \rightarrow D^* \tau^+ \nu_\tau)}{\text{BR}(B^+ \rightarrow D^* \ell^+ \nu_\ell)}$ \cite{RD}.
Observables in the Kaon sector include:
the branching ratios of the processes
$K^+ \rightarrow \pi^+ \nu \bar{\nu}$ \cite{Artamonov:2008qb},
$K^0_L \rightarrow \pi^0 \nu \bar{\nu}$ \cite{Ahn:2009gb},
the mass difference $\Delta M_K$
between $K_L$ and $K_S$ \cite{Agashe:2014kda},
and the indirect CP-violation
$\epsilon_K$
in the $K - \bar{K}$ system \cite{Agashe:2014kda}.
When calculating these observables, we take into account theoretical uncertainties as well as those from CKM matrix, rare decays, and hadronic parameters.
The experimental limits of these constraints are as follows,
\begin{eqnarray}
122.1 \text{ GeV} \leq m_h \leq 128.1 \text{ GeV}	\, ,	&&	
	\label{cons_mh}	\\
\Delta a_\mu = a_\mu^\text{exp} - a_\mu^\text{SM}
			 = (24.9 \pm 8.7) \times 10^{-10} \, , &&
	\label{cons_amu}	\\
\Delta \rho < 8.8 \times 10^{-4}	\,	,	&&	
	\label{cons_rho}	\\
17.715 \text{ ps}^{-1} \leq \Delta M_s \leq 17.799 \text{ ps}^{-1}	\,	,	& [2\sigma]&
	\label{cons_Ms}	\\
0.504 \text{ ps}^{-1} \leq \Delta M_d \leq 0.516 \text{ ps}^{-1}	\,	,	& [2\sigma]&
	\label{cons_Md}	\\
0.70 \times 10^{-4} \leq \text{BR}(B^\pm \rightarrow \tau^\pm \nu_\tau) \leq 1.58 \times 10^{-4}	\,	,	& [2\sigma]&
	\label{cons_btaunu}	\\
2.99 \times 10^{-4} \leq \text{BR}(\bar{B}^0 \rightarrow X_s \gamma) \leq 3.87 \times 10^{-4}	\,	,	& [2\sigma]&
	\label{cons_bsg}	\\
1.7 \times 10^{-9} \leq \text{BR}(B^0_s \rightarrow \mu^+ \mu^-) \leq 4.5 \times 10^{-9}	\,	,	& [2\sigma]&
	\label{cons_bmumu}	\\
0.84 \times 10^{-6} \leq \text{BR}(\bar{B}^0 \rightarrow X_s \ell^+ \ell^-)_\text{low} \leq 2.32 \times 10^{-6}	\,	,	& [2\sigma]&
	\label{cons_bsmumuL}	\\
2.8 \times 10^{-7} \leq \text{BR}(\bar{B}^0 \rightarrow X_s \ell^+ \ell^-)_\text{high} \leq 6.8 \times 10^{-7}	\,	,	& [2\sigma]&
	\label{cons_bsmumuH}	\\
2.7 \times 10^{-6} \leq \text{BR}(b \rightarrow d \gamma) \leq 25.5 \times 10^{-6}	\,	,	& [2\sigma]&
	\label{cons_bdg}	\\
\text{BR}(B^0_d \rightarrow \mu^+ \mu^-) \leq 8.7 \times 10^{-10}	\,	,	& [3\sigma]&
	\label{cons_bdmumu}	\\
\text{BR}(B \rightarrow X_s \nu \bar{\nu}) < 6.4 \times 10^{-4}	\,	,	& [90\% \text{CL}]&
	\label{cons_bxsnunu}	\\
\text{BR}(B^+ \rightarrow K^+ \nu \bar{\nu}) < 1.6 \times 10^{-5}	\,	,	& [90\% \text{CL}]&
	\label{cons_bknunu}	\\
\text{BR}(B^0 \rightarrow K^{*0} \nu \bar{\nu}) < 5.5 \times 10^{-5}	\,	,	& [90\% \text{CL}]&
	\label{cons_bksnunu}	\\
0.299 \leq R_D \leq 0.495	\,	,	& [2\sigma]&
	\label{cons_rd}	\\
0.259 \leq R_{D^*} \leq 0.373	\,	,	& [3\sigma]&
	\label{cons_rds}	\\
\text{BR}(K^+ \rightarrow \pi^+ \nu \bar{\nu}) < 4.03 \times 10^{-10}	\,	,	& [2\sigma]&
	\label{cons_kpinunu}	\\
\text{BR}(K^0_L \rightarrow \pi^0 \nu \bar{\nu}) < 2.6 \times 10^{-8}	\,	,	& [90\% \text{CL}]&
	\label{cons_kpi0nunu}	\\
5.275 \times 10^{-3} \text{ ps}^{-1} \leq \Delta M_K \leq 5.311 \times 10^{-3} \text{ ps}^{-1}	\,	,	& [2\sigma]&
	\label{cons_deltamk}	\\
2.206 \times 10^{-3} \leq \epsilon_K \leq 2.250  \times 10^{-3}	\,	.	& [2\sigma]&
	\label{cons_epsk}
\end{eqnarray}

Various constraints from direct searches for new particles at colliders are relevant for the scenarios we consider.  
While scenarios with light sfermions are severely restricted by LEP,  the LSP can be light enough to contribute to the $Z_1$ invisible decay width, we impose the constraint
$\Delta \Gamma_{Z_1}< 0.5$~MeV~\cite{Freitas:2014hra}.

Searches for a heavy neutral gauge boson in the dilepton and dijet channels have been performed at the LHC  both at  8 TeV and 13 TeV. 
We use the most recent data on the dilepton final state corresponding to an integrated luminosity of 3.2 fb$^{-1}$ at $\sqrt{s} = 13$ TeV  \cite{Aaboud:2016cth}.
Here, the $Z_2$ mass limit is interpolated for each specific value of $\theta_{E_6}$.
Limits from the dijet resonance searches at the LHC are obtained with the method described in Ref.~
\cite{Fairbairn:2016iuf}
using a combination of ATLAS 
\cite{Aad:2014aqa,ATLAS:2015nsi}
and CMS \cite{Khachatryan:2015sja,Khachatryan:2015dcf}
dijet data at 8 TeV and 13 TeV. These constraints are included in micrOMEGAs4.3~\cite{Barducci:2016pcb}.
In the UMSSM, there are cases where the lightest chargino is long-lived, typically when the chargino is nearly pure wino or nearly pure higgsino.
For such points, we take into account the results from long-lived chargino searches at the Tevatron and the LHC.
To derive this constraint, the observed limits for the cross sections of long-lived chargino pair production at D0 \cite{Abazov:2012ina} experiment are employed in combination with
the observed limit for chargino pair production and neutralino-chargino production cross section at the ATLAS \cite{ATLAS:2014fka} experiment. We follow the procedure described in~\cite{Belanger:2015cra}.

In addition to the constraints from collider physics, we take into account those from cosmological observations.
The most recent measurement of the DM relic density by Planck experiment \cite{Ade:2015xua} reads
\begin{eqnarray}
\Omega_\text{CDM} h^2 &=& 0.1188 \pm 0.0010	\,	.
\label{cons_omega}
\end{eqnarray}
In the global parameter scan, we impose only an upper bound on the DM relic density of the LSP. 
Thus we implicitly assume that there could be an additional  DM candidate.

For  DM direct detection the LUX experiment sets the most severe constraint on the spin-independent (SI) cross section between a DM particle and nucleons
\cite{Akerib:2013tjd,Akerib:2015rjg,Akerib:2016vxi}, while  PICO-60  \cite{Amole:2015pla}  sets the best direct limit on the spin-dependent (SD) cross section on protons.
The SD cross section on protons is also constrained by IceCube \cite{Aartsen:2016exj} by observing the neutrino flux from DM captured in the Sun, this limit however depends on specific annihilation channels.  

The UMSSM model with  kinetic mixing was  implemented in LanHEP version 3.2.0 
\cite{Semenov:2008jy,Semenov:2010qt,Semenov:2014rea}
which produces the model files suitable for CalcHEP 
\cite{Belyaev:2012qa}.
The spectrum and all the DM observables are calculated using micrOMEGAs  version 4.3.1 
\cite{Belanger:2011rs,Belanger:2015cra,Belanger:2014vza, Barducci:2016pcb}
with the help of UMSSMTools~\cite{DaSilva:2013jga} adapted from NMSSMTools v5.0.2 routines \cite{Ellwanger:2005dv,Domingo:2015wyn}. The latter includes in particular all flavour physics observables.
For collider observables, we use a routine of micrOMEGAs to compute the $Z_2$ limits from LHC as well as the $Z_1$ invisible width.
An interface to HiggsBounds \cite{Bechtle:2013wla} allows to test the Higgs sector of the model with respect to $95\%$ CL exclusion limits from the LEP, Tevatron and LHC experiments.
Finally the points satisfying all the above collider constraints are analysed with SModelS 1.0.4 which decomposes the signal of any BSM model into simplified topologies in order to test it against LHC bounds \cite{Kraml:2013mwa,Kraml:2014sna}.

\subsection{Benchmark analysis}

We examine the effect of the kinetic mixing on the sparticle spectrum for a benchmark set of the UMSSM inputs.
The simplified UMSSM input parameters are taken to be:
the common gaugino masses
$M_1'= M_1 = M_2 = M_3 = M_G = 3$ TeV,
the common slepton and squark soft masses
$m^0_{\tilde{l}} = 1.1$ TeV and
$m^0_{\tilde{q}} = 3$ TeV respectively,
the $Z_2$ boson mass
$M_{Z_2} = 3.8$ TeV,
the common trilinear coupling
$A_0 = 3$ TeV,
the $\mu$-parameter
$\mu = 1035.5$ GeV,
the mixing angle between two $Z$-bosons
$\alpha_Z = -0.64 \times 10^{-4}$, and
the angle
$\theta_{E_6} = 1.4$. 
Note that these  values of $\mu, \alpha_Z$, and 
$\theta_{E_6}$ are chosen randomly such  that all the phenomenological constraints, especially the 125 GeV Higgs boson mass and the DM relic abundance, can be satisfied for a suitable value of $k$.
Letting the kinetic mixing parameter to be a free input, we find that the range with $-0.742 < k < 0.399$ is theoretically acceptable.
The values outside this range are excluded by the reality condition (\ref{consb}) and the tachyonic slepton condition.

\begin{figure}
\centering
  \includegraphics[width=12cm]{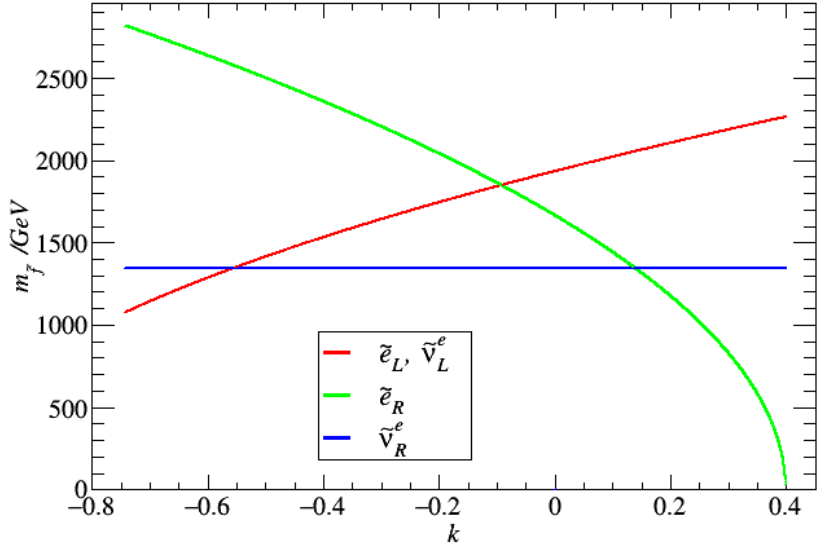}
\caption{Slepton masses as functions of the gauge kinetic mixing $k$.} 
\label{slepton_k}
\end{figure}

\begin{figure}
\centering
  \includegraphics[width=12cm]{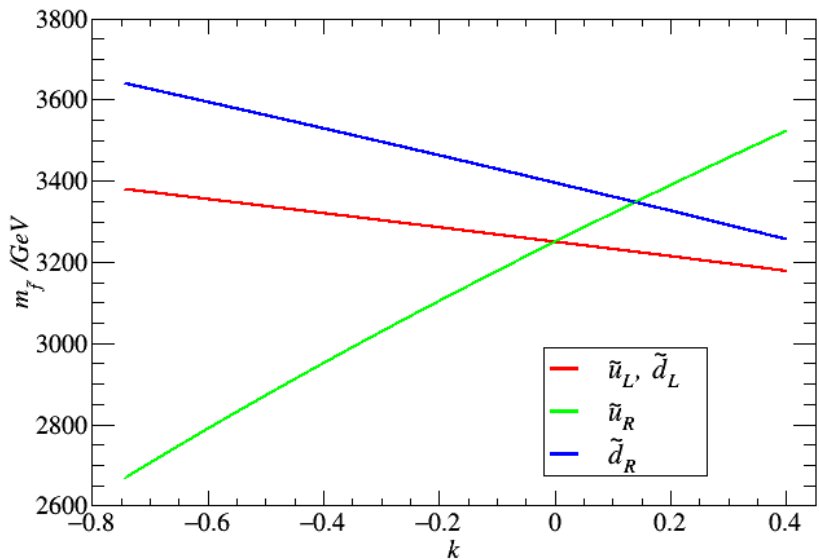}
\caption{Squark masses as functions of the gauge kinetic mixing $k$.} 
\label{squark_k}
\end{figure}

In Fig. \ref{slepton_k}, we show the dependence of slepton masses of the first generation on the kinetic mixing parameter. 
For this particular choice of inputs, the behaviors of the second and third slepton generations are very similar.
The two sfermions belonging to the LH slepton doublet,
$\tilde{\nu}^e_L$ and $\tilde{e}_L$, have masses too degenerate to be distinguished in the plot.
When increasing the kinetic mixing, the LH slepton masses increase while the RH selectron mass decrease, becoming  tachyonic for $k > 0.399$. The  RH sneutrino mass is nearly independent on $k$.

Fig. \ref{squark_k} shows the first generation squark masses as functions the kinetic mixing parameter. 
Here, only the RH up-squark becomes heavier for larger $k$.
The other squark masses (LH up-squark, LH down-squark, and RH down-squark) decrease with the kinetic mixing parameter $k$.
As for the slepton case, the other two generations of squarks have a similar behavior  as the first generation and are
therefore not shown in the figure.

The $k$-dependence of sfermion masses can be explained using Eqs. (\ref{D-term}), (\ref{gE-define}), and (\ref{Qp}).
Within the allowed range of $k$, corrections to the sfermion masses are dominantly controlled by $Q^p(k)$.
For the benchmark value $\theta_{E_6} = 1.4$, the quantity in the brackets of the right side of (\ref{D-term}) is positive.
Therefore, the D-term correction to a sfermion mass is approximately proportional to $\frac{k}{1-k^2}$ and its  hypercharge $Y$.
The dependence on $k$ is therefore stronger for  the sparticle with a large hypercharge $Y$. The mass increases (decreases) with $k$ for negative (positive) $Y$.
The RH sneutrino has a hypercharge $Y_{\tilde{\nu}_R}=0$, hence its mass  remains almost  constant.
The kinetic mixing enters the neutralino masses only through the mixing between higgsinos, singlino and bino', hence the neutralino masses are  almost  independent of the kinetic mixing.


\begin{figure}
\centering
  \includegraphics[width=12cm]{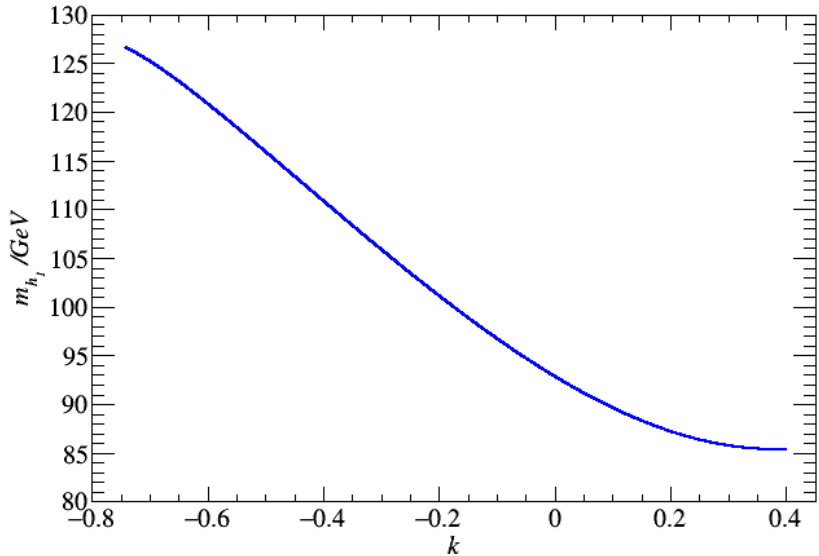}
\caption{The SM-like Higgs boson mass as a function of the gauge kinetic mixing $k$.} 
\label{mh_k}
\end{figure}

The SM-like Higgs boson mass is plotted as a function of the kinetic mixing $k$ in Figs (\ref{mh_k}).
We see that the Higgs boson mass decreases with $k$ and that in the absence of kinetic mixing the mass would be much below the observed value. Thus, enabling a negative nonzero kinetic mixing, in this case $k\approx -0.7$, allows to bring the Higgs boson mass  in agreement with the observed value, $m_h \sim 125$ GeV.


For illustration, we show in Tables \ref{benchmark1}, the sparticle mass spectrum as well as the constrained observables for the  benchmark just discussed. 
Assuming theoretical uncertainties in the calculations, we find that only the muon g-2 and $R_{D^*}$ satisfy the corresponding constraints at $3 \sigma$ level, while all other observables comply with the experimental limits at $2 \sigma$ level.
The LEP limits, the invisible $Z_1$ width, and the dilepton and dijet constraints for the $Z_2$ boson from the LHC, the constraint from long-lived chargino searches at D0 and the ATLAS experiments are all satisfied.
This benchmark is also compatible with limits on the Higgs sector obtained  by Lilith and  HiggsBounds as well as with limits on sparticles obtained with  SModelS.
We note that the kinetic mixing induces large shifts in the heavy Higgs doublet, from $\simeq 2.5$~TeV when $k=0$  to $\simeq 5$~TeV when $k=-0.7$ while the singlet mass, $m_{h_2}$ in Table 2, remains constant. Such heavy masses are in any case out of reach of the LHC.
The DM candidate for this benchmark is a higgsino-like neutralino.
Its relic density is achieved by annihilation into gauge bosons and  coannihilation  with the second lightest neutralino and the chargino NLSP whose masses are almost degenerate.
The SI and SD cross sections of the DM scattering on nuclei meet the requirement from the LUX and IceCube experiments.
Since the sparticles are quite heavy, it is challenging to test this benchmark at the LHC.
However, the future XENON1T will be able to test the model via the SI interaction of the neutralino DM.

\begin{table}
\begin{center}
\begin{math}
\begin{array}{ |cc||cc||cc| } 
\hline
\multicolumn{2}{|c||}{\text{Inputs}} & 
	\multicolumn{2}{c||}{\text{Mass spectrum}} & 
		\multicolumn{2}{c|}{\text{Observables}} \\ [0.5ex]
\hline \hline
M_G				& 3000 	& h_1		& 125.2 	& 
		\Delta a_\mu   &  3.182 \times 10^{-11} \\
		
m^0_{\tilde{l}}	& 1100 	& h_2		& 3800 	& 
		\Delta \rho  &  1.616 \times 10^{-6} \\ 
		
m^0_{\tilde{q}}	& 3000	& h_3		& 5024 	& 
		\Delta M_s   &  16.83 \text{ ps}^{-1} \\ 
		
M_{Z_2}			& 3800 	& A^0		& 5024  & 
		\Delta M_d   &  0.485 \text{ ps}^{-1} \\ 
		
A_0				& 3000  & H^\pm		& 5025   & 
		\text{BR}(B^\pm \rightarrow \tau^\pm + \nu_\tau)   & 1.070  \times 10^{-4} \\

\mu				& 1035.5	& Z_2		& 3800 	  & 
		\text{BR}(\bar{B}^0 \rightarrow X_s \gamma )  &  3.347 \times 10^{-4} \\
		
\alpha_Z	& -0.64 \times 10^{-4} 	& \tilde{g}	& 3000 & 
		\text{BR}(\bar{B}_s^0 \rightarrow \mu^+ \mu^-)  &  3.346 \times 10^{-9} \\
		
\theta_{E_6}	& 1.4 	& \tilde{\chi}^0_{1,2}	& 1033, 1036    & 
		\text{BR}(\bar{B}^0 \rightarrow X_s \ell^+ \ell^-)_\text{low} &  1.671 \times 10^{-6}\\
		
k	& -0.7  & \tilde{\chi}^0_{3,4}	& 2586, 3000    & 
		\text{BR}(\bar{B}^0 \rightarrow X_s \ell^+ \ell^-)_\text{high}  &  2.401 \times 10^{-7} \\
		
	&   & \tilde{\chi}^0_{5,6}	& 3003, 5586  &
\text{BR}(b \rightarrow d \gamma)  & 1.73 \times 10^{-5}  \\
		
	&  	& \tilde{\chi}^\pm_{1,2}& 1034, 3003    & \text{BR}(B^0_d \rightarrow \mu^+\mu^-)  &  9.62  \times 10^{-11}\\

	&  	& \tilde{\nu}^{e,\mu}_{L,R}		& 1141, 1345 & 
\text{BR}(B \rightarrow X_s \nu \bar{\nu})   &  2.89  \times 10^{-5} \\
		
	&	& \tilde{e},\tilde{\mu}_{L,R}	& 1144, 2764   & \text{BR}(B^+ \rightarrow K^+ \nu \bar{\nu})  & 3.96 \times 10^{-6} \\
	
	&	& \tilde{\nu}^\tau_{L,R}& 1141, 1100  & \text{BR}(B^0 \rightarrow K^{*0} \nu \bar{\nu})  & 9.15 \times 10^{-6} \\
	
	&	& \tilde{\tau}_{1,2}	& 1144, 2764   & R_D  & 0.297 \\
	
	&	& \tilde{u},\tilde{c}_{L,R}		& 3372, 2705 &   R_{D^*} & 0.252 \\
	
	&	& \tilde{d},\tilde{s}_{L,R}		& 3373, 3626  & \text{BR}(K^+ \rightarrow \pi^+ \nu \bar{\nu})  & 8.60 \times 10^{-11} \\
	
	&	& \tilde{t}_{1,2}		& 2700, 3385  &  \text{BR}(K^0_L \rightarrow \pi^0 \nu \bar{\nu}) & 2.70 \times 10^{-11} \\
	
	&	& \tilde{b}_{1,2}		& 3374, 3626  & 
\Delta M_K  & 5.80 \times 10^{-3} \\

	&	&	&	& \epsilon_K	& 1.86 \times 10^{-3} 	\\
	
	&	&	&	& \Omega h^2	& 0.1188 	\\

	&	&	&	& \sigma_{SI}^{\chi-p}	& 2.727 \times 10^{-10} 	\\

	&	&	&	& \sigma_{SD}^{\chi-p}	& 1.840 \times 10^{-7} 	\\
\hline
\end{array}
\end{math}
\caption{The sparticle mass spectrum in GeV and the corresponding constrained observables for the given set of input parameters.}
\label{benchmark1}
\end{center}
\end{table}

\subsection{Global parameter scan}

We assume that squark and slepton soft masses of the first two generations are universal, 
$m_{\tilde{f}_1} = m_{\tilde{f}_2}$ where $f = \{ q,u,d,l,\nu,e \}$,
and the trilinear couplings of the first two generations are negligible.
We are thus left  with 25 free parameters including the gauge kinetic mixing.
The ranges for these parameters are chosen as follows 
\begin{eqnarray}
&	-2000  < M_1, M_2, \mu < 2000 \text{ (GeV)}	&	\\
&	100  < M_3 < 4000 \text{ (GeV)} ,	&	\\
&	10< M'_1 < 4000 \text{ (GeV)} ,	&	\\
&	0 < m_{\tilde{q}_i}, m_{\tilde{u}_i} , m_{\tilde{d}_i} < 4000 \text{ (GeV)} , &
	\quad i = \{1,2,3\}		\\
&	0 < m_{\tilde{l}_i}, m_{\tilde{\nu}_j}, 
	m_{\tilde{\nu}_{\tau R}}, m_{\tilde{e}_i} < 2000 \text{ (GeV)} ,	&	\quad
	i = \{1,2,3\}, \, j = \{1,2\}	\\
&	-4000  < A_\lambda, A_t, A_b, A_\tau < 4000 \text{ (GeV)} ,	&	\\
&	-10^{-3} < \alpha_Z < 10^{-3} ,	&	\\
&	1000 < M_{Z_2} < 8000 \text{ (GeV) },	&	\\
&	-\frac{\pi}{2} < \theta_{E_6} < \frac{\pi}{2},	&	\\
&	-1 < k < 1.	&
\end{eqnarray}
We have performed a random scan over this parameter region with $5\times 10^8$ points.
The particle mass spectrum and all the above constrained observables have been calculated for each point.
Only points satisfying the theoretical constraints are considered in  our analysis.


\begin{figure}
\centering
  \includegraphics[width=12cm]{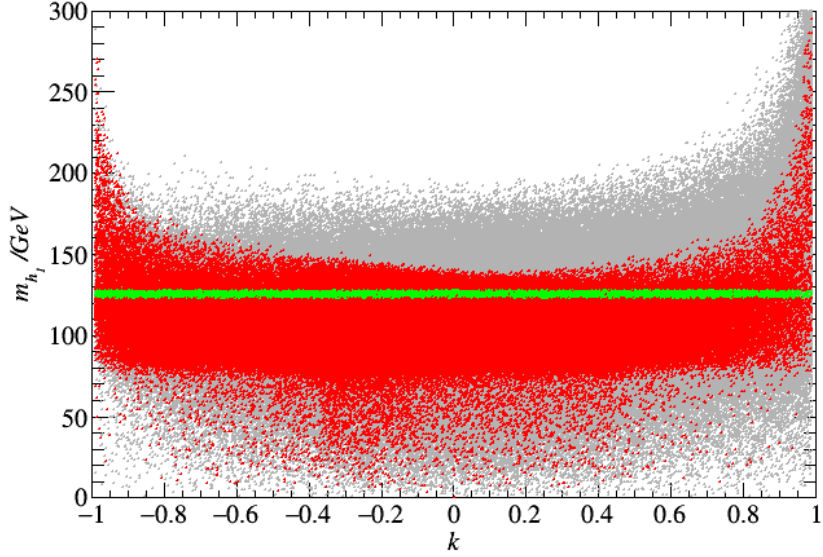}
\caption{Theoretically allowed points in the plane $(k, m_{h_1})$. Grey points are excluded by flavour constraints, LEP mass limits, invisible $Z_1$ decay width,  LHC searches for a $Z_2$ boson, and  searches for long-lived chargino, red points are excluded by Higgs mass constraints as well as HiggsBounds and Lilith. Green points satisfy all constraints.
} 
\label{plot_k_mh}
\end{figure}

\begin{figure}
\centering
  \includegraphics[width=12cm]{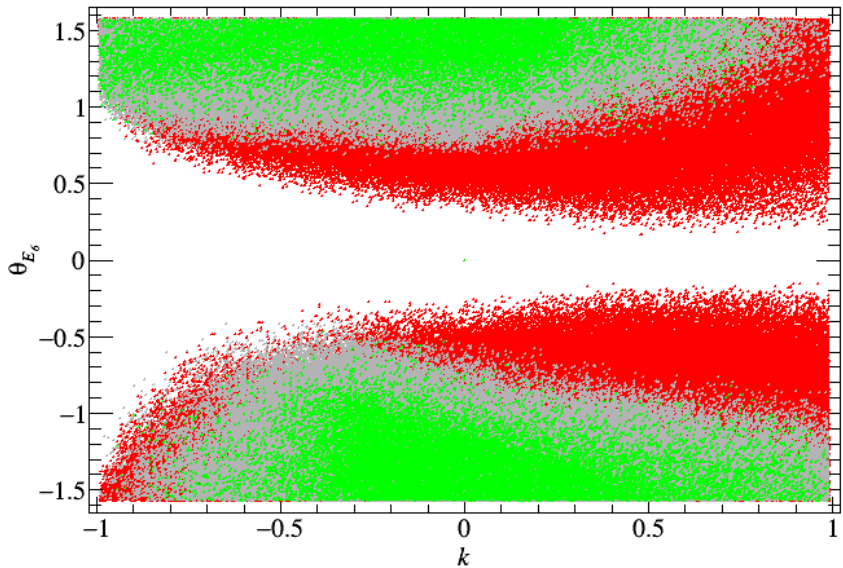}
\caption{Allowed region in the plane $k, \theta_{E_6}$ (in green). Red points are excluded by $Z_2$ searches while grey points are excluded by all other constraints.
} 
\label{plot_k_tE6}
\end{figure}

In Fig.~\ref{plot_k_mh}, a scatter plot is shown in the plane $(k,m_{h_1})$.
Looking at the density in the plot, most of the points correspond to a Higgs mass between  50 GeV and 180 GeV.
Clearly the Higgs mass in this model can be enhanced with the nonzero gauge kinetic mixing and 
the effect of nonzero kinetic mixing can be quite large for $|k| \in (0.8,1.0)$.
This  is due to the $U(1)'$ D-term contribution (the third term in Eq. (\ref{mh1})) proportional to the gauge coupling $g_E$ that is 
greatly enhanced for large $k$.
Nevertheless, the value  $m_{h_1} \simeq 125$~GeV can be obtained for any value  of the kinetic mixing.
Note that the constraints from flavour physics (\ref{cons_rho})-(\ref{cons_epsk}) and  $(g-2)_\mu$ (\ref{cons_amu})
 are most  severe in the low Higgs mass region  $m_{h_1} \lesssim 80$ GeV and relatively small kinetic mixing, a region that is in any case not physically interesting. 
For $m_{h_1} \simeq 125$ GeV, the  constraint  from $Z_2$ searches plays a very important role while  the searches for long-lived chargino do not provide a significant constraint after all other phenomenological constraints are taken into account.
The impact of the $Z_2$ searches and other collider constraints is best illustrated  in the plane $(k,\theta_{E_6})$, see Fig.~\ref{plot_k_tE6}.
The $Z_2$ constraint is particularly important  in the  region with large kinetic mixing $k \gtrsim 0.5$ due to the enhancement of the coupling $g_E$,
it has also an impact for large negative values of the kinetic mixing although for these values  theoretical constraints are more important and only allow a restricted range for $\theta_{E_6}$.

\begin{figure}
\centering
  \includegraphics[width=12cm]{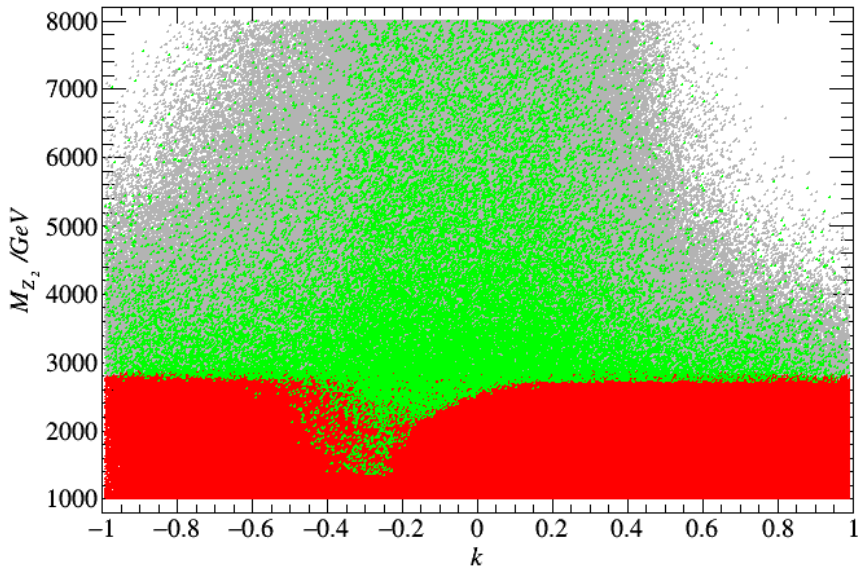}
\caption{Allowed region in the plane  $k, M_{Z_2}$ (in green).
The color code is the same as in Fig. \ref{plot_k_tE6}.
} 
\label{plot_k_mz2}
\end{figure}

It is interesting to observe that because of the non zero kinetic mixing, the LHC constraints on the $Z_2$ mass can be significantly relaxed. Typically the current limit is around 2.8~TeV with a slight dependence on $\theta_{E_6} $. However for certain values of $\theta_{E_6}$ and $k$ the coupling of the $Z_2$ to the RH leptons is strongly suppressed due to a cancellation between the two terms in Eq.~\ref{Qp}. Hence without an  increase in the coupling to LH leptons or quarks to compensate, the limit on  the $Z_2$ boson mass from dileptons is relaxed and can drop below 2~TeV.
 This occurs for $-1.2<\theta_{E_6}<-0.8$ and $-0.45< k<-0.15$, see Fig.~\ref{plot_k_mz2}.  The lightest allowed value  is found  to be $M_{Z_2}=1.3$~TeV.


\begin{figure}
\centering
  \includegraphics[width=12cm]{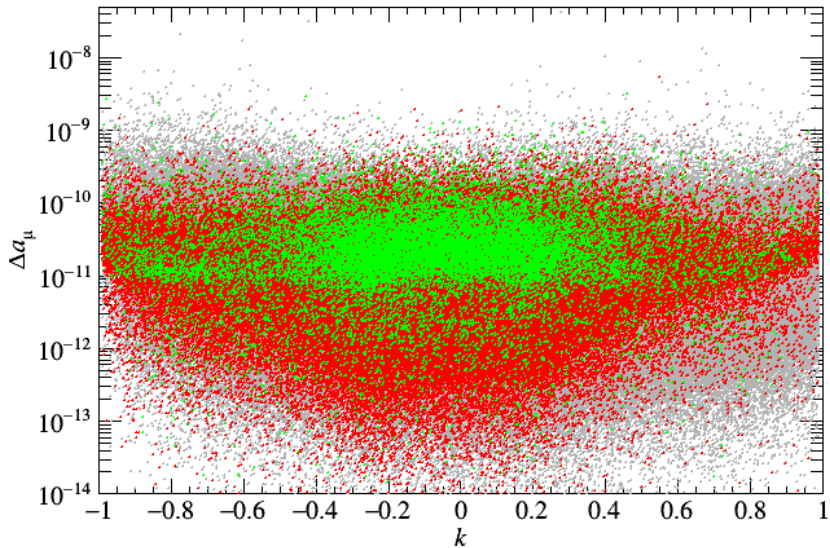}
\caption{The anomalous magnetic moment of the muon, $\Delta a_\mu$  vs the gauge kinetic mixing $k$. Green points are allowed, red points are excluded by constraints on the Higgs mass and couplings and grey points are excluded by all other flavour and collider constraints.
The green points are plotted on top of the other ones.
} 
\label{plot_k_Damu}
\end{figure}

The predictions for the muon anomalous magnetic moment are presented  in Fig. \ref{plot_k_Damu}. It is possible, for any value of $k$ to reproduce the 
central value for for this observable, Eq.~\ref{cons_amu}. In particular $\Delta a_\mu$  can  be enhanced by a relatively light $Z_2$,
this is because a small $M_{Z_2}$ implies a small Higgs singlet VEV, $v_S$, (see Eq. (\ref{mz2})) which in turn  can result in a small $U(1)'$ D-term corrections to smuon masses (Eq.~\ref{D-term}).
However after taking into account  the LHC constraints on the $Z_2$ boson,  the number of points with $\Delta a_\mu\approx 10^{-9}$ is significantly reduced especially at large $k$. In addition the region of parameter space where the $Z_2$ mass can be relaxed does not correspond to the one leading to a large contribution to $\Delta a_\mu$.
Other constraints also restrict the allowed regions. In the flavour sector (also shown in grey in Fig.~\ref{plot_k_Damu}),
the constraints on $\Delta M_s$ (\ref{cons_Ms}) and $\Delta M_d$ (\ref{cons_Md}) are quite important but mostly  for the region with low  $\Delta a_\mu < 10^{-11}$.
while for $\Delta a_\mu \gtrsim 10^{-10}$,  the constraints from  $\bar{B}^0 \rightarrow X_s \gamma$ (\ref{cons_bsg}) are more severe. Note that 
the constraints for the branching ratio of $B_d^0 \rightarrow \mu^+ \mu^-$ and $R_{D^*}$ are particularly severe and  that we have used the $3\sigma$ bounds for them.
As mentioned above,  the constraints on the Higgs sector (in red) have a strong impact on the parameter space, they rule out a large number of points with low $\Delta a_\mu$, but also some points at large $k$ where the prediction for $\Delta a_\mu$ is within the $2\sigma$ observed range. 
It is worth noting that within our scan, there are no green points with $\Delta a_\mu \gtrsim 2 \times 10^{-9}$ in the region $|k| < 0.2$, while
we can find such green points for the regions $|k| > 0.2$.
Therefore, even after taking into account all constraints, there is some enhancement in the predicted value of the muon $g-2$ although it is small.


\begin{figure}
\centering
  \includegraphics[width=12cm]{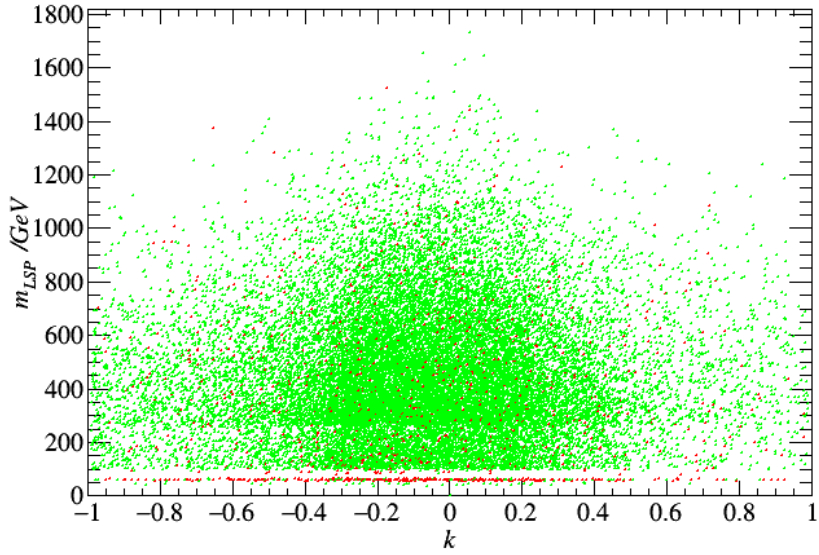}
\caption{The scatter plot on the plane $(k, m_{\rm LSP})$ for the neutralino-LSP (green) and the RH sneutrino-LSP (red).
All the points satisfy the collider and flavour constraints as well as the upper bound on the relic density.} 
\label{plot_k_mLSP}
\end{figure}

We now discuss dark matter observables including the relic density and DM elastic scattering cross section on nuclei. In the numerical results we consider only points that successfully  predict a Higgs boson mass in the range $[122,128]$ GeV and that satisfy collider and flavour constraints. Points with a charged/colored LSP are  not considered as they are disfavored by cosmological observations.
There are two posibilities for DM in this model, the lightest neutralino or the RH sneutrino,  recall that the LH sneutrino LSP typically leads to
a large SI cross section on nuclei  due to the exchange of a  $Z_1$ boson and are thus excluded by DM direct detection experiments \cite{Falk:1994es}. We do not consider this possibility.
Fig. \ref{plot_k_mLSP} shows the scatter plot of the collider allowed points in  the plane $(k, m_{\rm LSP})$, including both
the neutralino-LSP (green) as well as the  RH sneutrino-LSP (red). The latter is much less likely. Moreover we observe that in the region with small $|k|$ there are more possibilities to have a heavy DM than in the region with large $|k|$.
In particular most of the points associated with a DM candidate with a mass $m_{\rm LSP} \ge 1200$~GeV have relatively small kinetic mixing $|k| \le 0.5$. 
This is related to the fact that large kinetic mixing induces a  shift in some of the sfermion masses and are therefore more likely to have a charged LSP and moreover
 that more points are excluded by the LHC search for $Z_2$ boson due to its enhanced coupling for large $|k|$, see Fig. \ref{plot_k_tE6}.
 Note that Fig.~\ref{plot_k_mLSP} shows only the points that satisfy the relic density upper bound from PLANCK, however  the distribution of points
is similar without the constraint except for the region with a very light LSP.

\begin{figure}
\centering
  \includegraphics[width=12cm]{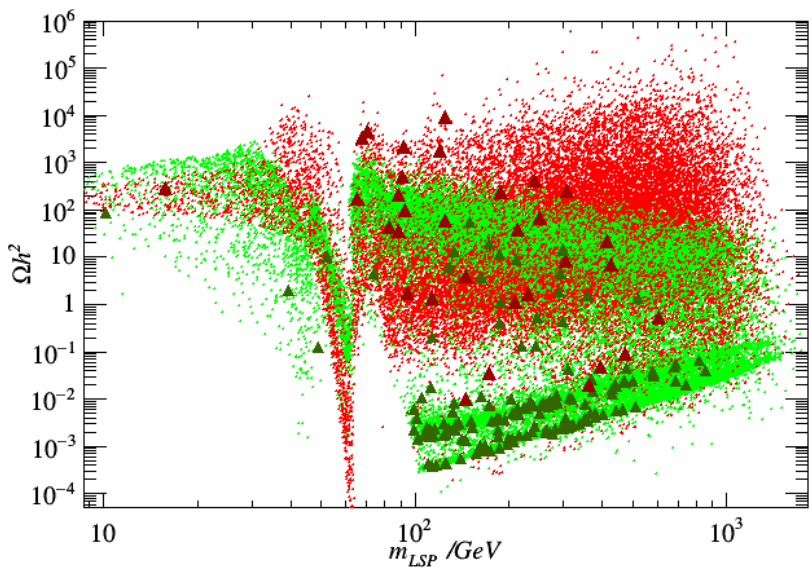}
\caption{The scatter plot on the plane $(m_{\rm LSP}, \Omega h^2)$.
The color codes are the same as in Fig. \ref{plot_k_mLSP}.
Triangles: points satisfying the muon $g-2$ limits at $2\sigma$ level and SModelS test in addition to other collider constraints.} 
\label{plot_mLSP_Omega}
\end{figure}

%
The DM relic density for each type of DM is presented in the scatter plot of Fig. \ref{plot_mLSP_Omega}. 
We find as expected that the RH sneutrino DM is typically overabundant since it is very weekly coupled to SM particles. 
There are however special cases where the RH sneutrino predicts a relic density in agreement or below the PLANCK value. When the RH sneutrino mass  is near $m_{h_1}/2$, the DM annihilation is enhanced by a resonance effect, 
similarly when the mass is near  $m_{Z_2}/2$. Considering the LHC constraint on the $Z_2$, this requires a rather heavy sneutrino DM. 
Finally there is always the possibility of coannihilation with other sfermions or neutralino/chargino. The latter can occur for any mass. Although there is an impact of the gauge kinetic mixing in these scenarios since the masses of the Higgs, $Z_2$ and sfermions all depend on the kinetic mixing, in the global scan there is no direct correlation between the relic density and the kinetic mixing, and a value $\Omega h^2 \sim 0.1$ can be obtained for the whole range of $k$. 
This statement also holds for neutralino DM. 

The value of the relic density  for neutralino DM (in green in Fig.~\ref{plot_mLSP_Omega}) features a strong dependence on the nature of the LSP. 
Most of the points with overabundant DM are associated with a bino or singlino LSP while those with underabundant DM correspond to higgsino-like and wino-like LSP. These are clustered in the two strips at the bottom of  Fig. \ref{plot_mLSP_Omega} that extend from 100 GeV to 2 TeV, the wino-like LSP corresponding to the points with the lower value of the relic density.
Note that we find only a few points with $\tilde{S}$-like LSP, and no point with $\tilde{B}'$-like LSP.
Other green points predicting the allowed DM relic density are either well-tempered neutralino or bino-like neutralinos with co-annihilation. They scatter in a large mass range.
For $m_{\rm LSP} \lesssim 100$ GeV, and $\Omega h^2 \lesssim 0.1$, we find that the only possibility for the neutralino LSP to have an acceptable relic density is the Higgs-resonance region where the DM mass is about half the SM-like Higgs or $Z_1$ boson mass, $m_{\rm LSP} \sim 62$ GeV or  $m_{\rm LSP} \sim 45$ GeV.
Here, neutralino annihilation mainly happens via the Higgs ($Z_1$) exchange in the s-channel.
Altogether most of the points which satisfy the relic density upper bound are associated with a compressed spectra, in particular with a NLSP nearly degenerate with the LSP. Indeed such spectra is found for dominantly higgsino and wino neutralino as well as for coannihilation of neutralino or sneutrino. Because of the strong constraints from LEP on light charged particles below 100 GeV, coannihilation in this region is not possible. In addition several points (especially for sneutrino DM) are clustered around $m_{h_1}/2$, only a few points feature uncorrelated a large mass splitting between the NLSP and the LSP. 

\begin{figure}
\centering
  \includegraphics[width=12cm]{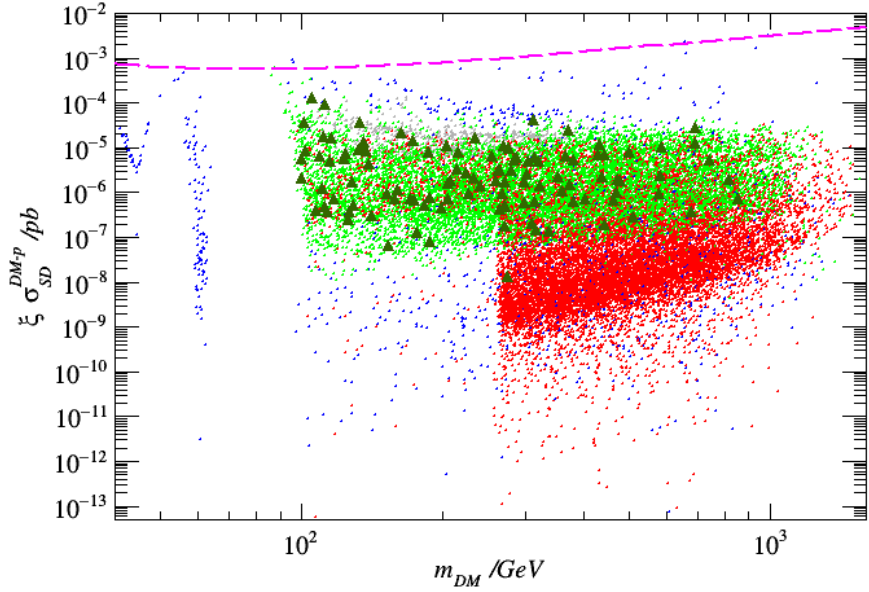}
\caption{The scatter plot on the plane $(m_{DM}, \sigma_\text{SD}^\text{DM-p})$.
Grey points are excluded by the IceCube constraint.
Other colored points satisfy all collider constraints, as well as the upper limit of DM relic density from Planck and the upper limit of SD cross section from IceCube.
Blue: bino-like LSP,
red: wino-like LSP,
green: higgsino-like LSP.
Triangles are a subset of green points, and satisfy the muon $g-2$ limits at $2\sigma$ level as well as SModelS test.
Dashed pink curve: PICO upper limit.} 
\label{plot_k_sigmaSD}
\end{figure}

The SD cross section for DM scattering on nuclei relevant to direct DM searches and to indirect searches with neutrino telescopes is shown in Fig. (\ref{plot_k_sigmaSD}) for all the points satisfying the collider constraints and the upper limit of DM relic density from Planck ($\Omega h^2 < 0.1208$).
Note that this figure includes only the neutralino-LSP  since the sneutrino-LSP is a scalar particle and does not have SD interaction with nuclei.
To take into account the fact that neutralino DM may account only for a fraction of the DM content of the universe, we rescale the cross section for cases where DM is underabundant by $\xi = \Omega h^2/0.1188$.
In the figure, points satisfying the muon $g-2$ at $2 \sigma$ level and all other collider constraints including LHC limits from SModelS  are marked by triangles. 
Moreover, we impose the IceCube limit as described in Ref~\cite{Belanger:2015hra}, the points  ruled out by this constraint  are shown as grey dots.
The  direct detection limit from PICO~\cite{Amole:2015pla} is also displayed as a  pink dashed line, this limit is easily satisfied for most of the points since they
 lie generally two orders of magnitude below the current limit. %
The sharp cut in the plot around 250 GeV is due to the constraint from searches for long-lived chargino.

\begin{figure}
\centering
  \includegraphics[width=12cm]{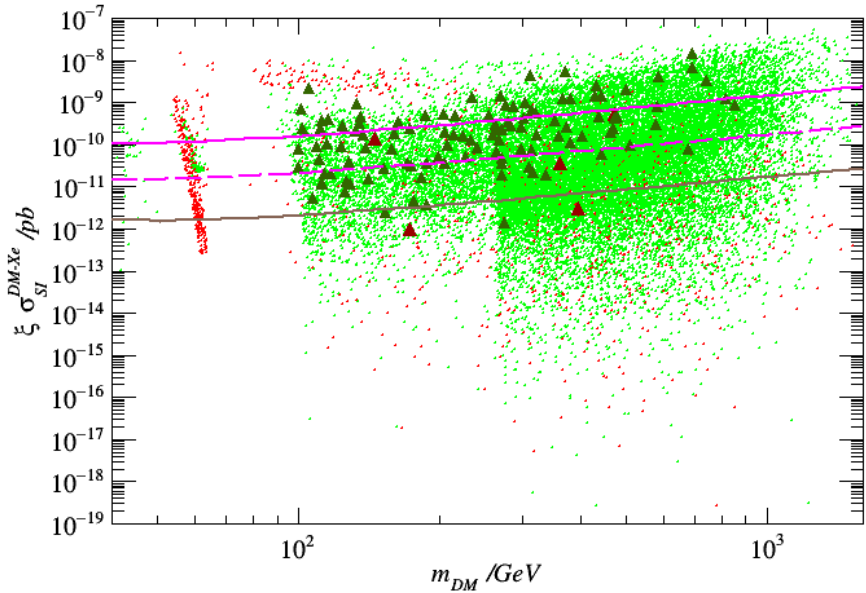}
\caption{The scatter plot on the plane $(m_{DM}, \sigma_\text{SI}^\text{DM-Xe})$.
All the points satisfy the collider constraints, as well as the upper limit of DM relic density from Planck and the upper limit of SD cross section from IceCube.
The color codes are the same as in Fig. \ref{plot_k_mLSP}.
Among these points, the triangles represent points satisfying the muon $g-2$ limits at $2\sigma$ level as well as SModelS test.
The pink, dashed pink, and brown curves indicate the upper limit set by 
LUX expriment \cite{Akerib:2016vxi}, the projected XENON1T and XENONnT \cite{Aprile:2015uzo} respectively.
} 
\label{plot_k_sigmaSI}
\end{figure}

Fig. \ref{plot_k_sigmaSI} shows the rescaled  SI cross section with respect to the DM mass.
The color codes are the same as in Fig. \ref{plot_k_mLSP}.
When the LSP is the RH sneutrino, a significant fraction of the scenarios are clustered around $m_{\rm LSP} \approx m_{h_1}/2$ in order to benefit from resonance annihilation in the early universe. The points
 that are further away from the resonance and that  require a larger coupling to the Higgs  lead to a large direct detection cross section and are excluded by LUX. Other points where the coupling to the Higgs is small predict a SI cross section that can be below the expected limit from XENONnT.  Other scenarios with a RH sneutrino are  ruled out by the LUX constraint, typically those that have a large coupling to the Higgs 
 while others that rely on co-annihilation predict a cross section orders of magnitude below the sensitivity of ton-scale experiments. 
The LUX limit  excludes a considerable
 number of points with the neutralino-LSP (green dots).
Especially, the LUX limit rules out several of the points that are compatible with muon $g-2$ at $2\sigma$ level (triangles).
In fact, with the expected sensitivity of the ton-scale experiments (for example the  projected sensitivity of XENON1T  as represented by the pink dashed curve), if no DM signal is observed 
 it will introduce a severe tension for the model to reconcile both the muon anomalous magnetic moment 
and the SI cross section between the neutralino-LSP and nuclei while  satisfying all other constraints. 
However, if we relax the muon $g-2$ constraint to 3$\sigma$ bounds, 
there are several neutralino-LSP scenarios, as in the MSSM, which can escape the most sensitive future direct detection experiments such as XENONnT 
\cite{Aprile:2015uzo}, typically they are associated with wino or higgsino LSP. 

%

\section{Conclusions}

%
We have implemented the gauge kinetic mixing in the UMSSM and
have found that the kinetic mixing has an important effect on the mass spectrum and on the coupling of the $Z_2$ boson with fermions.
Especially, the mass of the SM-like Higgs boson can be significantly enhanced by an appropriate choice of the kinetic mixing.
This then impacts physical observables and has been illustrated for a  benchmark point. 
After applying various 
 theoretical and phenomenological constraints and performing a global parameter scan we have shown
that for specific values of the kinetic mixing it is possible to 
 relax the current constraint on the $Z_2$ boson mass ($M_{Z_2} \lesssim 2.8$ TeV) to as low as  1.3 TeV when  its dilepton branching ratio is suppressed.
We have also found that  the predictions for $\Delta a_\mu$ in scenarios with gauge kinetic mixing can  in some cases be within $2\sigma$ of the measured value, thus releasing some of the tension on this observable. The properties of the LSP and the DM relic density have been examined, and it is found that agreement with the observed value of the relic density could be obtained for any value of the kinetic mixing. Both neutralino and RH sneutrino can be viable DM candidates.  Moreover direct DM searches play an important role in ruling out large portions of the parameter space and offer good prospects of probing the model further although not completely.  This is a feature shared with the MSSM. The RH sneutrino in particular can lead to very small scattering cross sections on nuclei.

The upcoming LHC runs with improved reach for the search of a  $Z_2$ boson will provide further decisive tests of the model, both in the dilepton and dijet modes. The latter being crucial to probe the cases where the branching ratios of $Z_2$ into dileptons is suppressed  because of the kinetic mixing. 
SUSY searches are more challenging since the model often features a compressed spectra, although searches for long-lived charged particles will probe a fraction of the compressed scenarios.

\section*{Acknowledgment}

The authors are grateful to Alexander Pukhov and Ursula Laa for useful discussions.
H.M.T. would like to thank LAPTh, especially  Patrick Aurenche, for  hospitality and support during his visit.
This work  is  supported  by  the  ``Investissements  d'avenir,  Labex   ENIGMASS"  and  by  the  French   ANR,   Project
DMAstro-LHC,   ANR-12-BS05-006.
The work of H.M.T. is partly supported by Vietnam National Foundation for Science and Technology Development (NAFOSTED) under the grant No. 103.01-2014.22.


\end{document}